\documentclass[journal,draftclsnofoot,onecolumn,12pt]{IEEEtran}

\usepackage[T1]{fontenc}% optional T1 font encoding

\usepackage{cite}
\usepackage[colorlinks,
linkcolor=red,
anchorcolor=blue,
citecolor=green
]{hyperref}
\ifCLASSINFOpdf
\else
\fi
\usepackage{amsmath}
\usepackage{diagbox}

\usepackage{ amssymb }
\usepackage{amsfonts}
\usepackage{caption}
\usepackage{graphicx}
\usepackage{lettrine}
\usepackage{epstopdf}
\usepackage{textcomp,booktabs}
\usepackage[usenames,dvipsnames]{color}
\usepackage{colortbl}
\definecolor{mygray}{gray}{.9}
\definecolor{mypink}{rgb}{.99,.91,.95}
\definecolor{mycyan}{cmyk}{.3,0,0,0}
\usepackage{pifont}
\usepackage{subfig}
\usepackage{multirow}
\usepackage{url}
\usepackage{color}
\usepackage{threeparttable}
\usepackage{setspace}
\usepackage{lineno}

\usepackage[dvipsnames,usenames]{color}

\usepackage{bm}

\usepackage{algorithm} %format of the algorithm
\usepackage{algorithmic} %format of the algorithm

\usepackage{dblfloatfix}

\usepackage[dvipsnames,usenames]{color}
\usepackage[usenames,dvipsnames]{color}

\definecolor{light-gray}{gray}{0.90}
\begin{document}
%\linenumbers
%\pagestyle{empty}  % no page number for the second and the later pages
%\topmargin=-0.6 truein
%\leftmargin=-0.6 truein
%\rightmargin=-0.6 truein
%\textheight=9.15 truein
%\oddsidemargin=-0.15truein
%\evensidemargin=-0.15truein
%\textwidth=6.5truein
	\title{Wireless Semantic Communications\\ for Video Conferencing}

	\author{Peiwen Jiang, Chao-Kai Wen, Shi Jin, and Geoffrey Ye Li
			\thanks{P. Jiang and S. Jin are with the National
				Mobile Communications Research Laboratory, Southeast University, Nanjing
				210096, China (e-mail: PeiwenJiang@seu.edu.cn; jinshi@seu.edu.cn).}
			\thanks{C.-K. Wen is with the Institute of Communications Engineering, National
				Sun Yat-sen University, Kaohsiung 80424, Taiwan (e-mail: chaokai.wen@mail.nsysu.edu.tw).}
			\thanks{G. Y. Li is with the Department of Electrical and Electronic Engineering,
				Imperial Colledge London, London, UK (e-mail: geoffrey.li@imperial.ac.uk).}}
	
	\maketitle
	\pagestyle{empty}  % no page number for the second and the later pages
	\thispagestyle{empty} % no page number for the first page
	%\vspace{-0.5cm}
	%\begin{abstract}
%		
%		
%		\setlength{\belowcaptionskip}{-0.3cm}   %调整图片标题与下文距离
%		\begin{figure*}[!t]
%			\centering
%			\includegraphics[width=6in]{gaoxuanxuan2}
%	
\begin{abstract}
		
Video conferencing has become a popular mode of meeting even if it consumes considerable communication resources. Conventional video compression causes resolution reduction under limited bandwidth.  Semantic video conferencing maintains high resolution by transmitting some keypoints to represent motions because the background is almost static, and the speakers do not change often. However, the study on the impact of the transmission errors on keypoints is limited. In this paper, we initially establish a basal semantic video conferencing (SVC) network, which dramatically reduces transmission resources while only losing detailed expressions. The transmission errors in SVC only lead to a changed expression, whereas those in the conventional methods destroy pixels directly. However, the conventional error detector, such as the cyclic redundancy check, cannot reflect the degree of expression changes. To overcome this issue, we develop an incremental redundancy hybrid automatic repeat-request (IR-HARQ) framework for the varying channels (SVC-HARQ) incorporating a novel semantic error detector. The SVC-HARQ has flexibility in bit consumption and achieves good performance. In addition, SVC-CSI is designed for channel state information (CSI) feedback to allocate the keypoint transmission and enhance the performance dramatically.  Simulation shows that the proposed wireless semantic communication system can significantly improve the transmission efficiency.

	\end{abstract}
	
	
	\newpage
	
	%\vspace{0.2cm}
	%\renewcommand{\baselinestretch}{1.5}
	\setlength{\baselineskip}{22pt}
	\section{Introduction}

 	\IEEEPARstart{I}{n} 
 semantic communications\cite{bao2011towards}, the shared  and local knowledge, extracted from the set of transmission contents,   helps  compress transmission information and correct the transmission errors according to semantic correlation.  However, the design of a practical semantic transceiver is difficult, especially the extraction of semantic knowledge using traditional methods due to lack of appropriate mathematical models.  Deep learning (DL) makes the implementation of semantic communication possible as evidenced by  many related works \cite{qin2021semantic,lu2021rethinking,sana2022learning}. 
 
 DL can potentially address many challenging issues \cite{FCDNN,ye2020deep,8054694,qin2018Deep,DBLP:journals/corr/abs-1809-06059} in communication systems, such as nonlinear interference,  insufficient pilots, channel estimation, channel coding,  channel state information (CSI) feedback, and autoencoder-based communication systems. These studies focus on the technical level and exploit DL to extract the features of channel environments   and outperform the traditional design. In semantic communications, DL is used to extract the semantic information from contents, and shared and local knowledge in the semantic level is implicitly contained in the trained parameters.  DL-based semantic communication systems are usually designed  using  joint source and channel coding methods and trained for a specific transmission content, including image \cite{bursalioglu2013joint,DJSCCC,bourtsoulatze2019deep,8731945}, video \cite{zhai2005joint},  speech \cite{weng2021semantic}, and text \cite{farsad2018deep,xie2020deep,9252948}.  Furthermore, for some specific tasks, such as object recognition \cite{8731945} and scene classification \cite{kang2021task} tasks, semantic communications can  significantly reduce the transmission overhead.

Video conferencing becomes an essential part of our work at present, especially during the pandemic of COVID-19. A high-resolution video transmission requires a  huge amount of transmission resources. Therefore, it is very challenging, especially for  conferencing over mobile phones. Driving a video with a few keypoints \cite{siarohin2019first} is widely applied in face-swapping. The keypoints of the frames from a driving video are extracted to represent the motion of facial expressions, and a generator is exploited to enable a source image to move similarly to the driving video. In \cite{wang2021one}, the facial driving  video is considered   the transmitted video, and the receiver can restore the driving video from the transmitted keypoints and the photo of the speaker. However, the existing studies only relate to the source coding module, and the impact of the errors from the physical channel transmission  is unclear. In addition, a  semantic-based video conferencing transmission should protect key information facing physical transmission errors and the received videos should be  acceptable under varying channels.

A basal network for the  semantic-based video conferencing called SVC is  established into a three-level framework. First, we investigate the entire transmission process and analyze the difference between the semantic and conventional errors. Then, we add acknowledgement (ACK) feedback to the SVC; it is a widely used technique in the conventional wireless communications to ensure a successful semantic transmission.  An incremental redundancy hybrid automatic repeat request (IR-HARQ) framework for the SVC, called SVC-HARQ, is proposed to guarantee the quality of the received frames when facing wicked channels. Then, the transmitter learns to allocate information with different importance according to signal-to-noise ratios (SNR) at different subchannels with the help of CSI, which is called SVC-CSI. The major contributions of this work are summarized as follows:

	1) \emph{Establishing SVC framework.}  The state-of-art technology to restore the image from several keypoints has achieved a huge compression ratio. Thus, we  exploit the technology  to cope with the  channel distortion.  Different from   the huge compression ratio (only few number of keypoints) that causes the transmitted image to lose the detailed expressions, the simulation results show that  transmission errors in physical channel transmission may change the locations of the keypoints and lead to inaccurate expressions. Nevertheless, these errors may be visually acceptable through semantic processing.  On the contrary, the errors usually destroy the pixels directly for the conventional methods.
	
	2) \emph{Combining with HARQ scheme.} To guarantee the feasibility of the SVC under varying channels,  the IR-HARQ feedback framework for the SVC, called SVC-HARQ, is developed. Compared with the conventional bit error detection  using cyclic redundancy check,  a semantic error detector is used to decide whether the received frame requires an incremental transmission. The semantic error detector exploits the fluency of the video to check of the received frame.  The simulation results demonstrate that the inaccurate keypoints usually  reduce the fluency. The proposed  SVC-HARQ can adapt different bit error rates (BER) and require to transmit fewer bits than the competing methods.

	3) \emph{Exploiting CSI.}  The  CSI is exploited so that optimal transmitted information with different importance can be allocated automatically on different subchannels, which is called SVC-CSI. The SVC-CSI learns to allocate more information at the subchannels with high SNRs  than those with low SNRs. An extra incremental transmission is trained without employing CSI because the performance of the SVC-CSI becomes worse when the testing channel environment is different from the training environment;    thus, it shall be robust to  varying channels and is called SVC-CSI-HARQ.

	The rest of this paper is organized as follows. Section \uppercase\expandafter{\romannumeral2} introduces the system model and the related methods, including conventional IR-HARQ and adaptive modulation. Then, we describe the proposed networks  in Section \uppercase\expandafter{\romannumeral3}. In Section \uppercase\expandafter{\romannumeral4}, we demonstrate the superiority of the proposed networks in terms of semantic metrics and the required  bits. Finally,  Section \uppercase\expandafter{\romannumeral5} concludes this paper.
	
\setlength{\baselineskip}{20pt}	
	\section{System Model and Related Works}
In this section, we first describe the existing frameworks on  semantic networks and then introduce some important techniques in wireless communications that can potentially help semantic transmission. Finally, we discuss the challenges when applying the semantic transmission  over the wireless communication systems.
\subsection{Semantic Frameworks}
To transmit  source information, such as a picture $\mathbf{p}$,  the semantic transmitter first extracts its meaning. The semantic extractor plays a similar role to the source encoder in the conventional communication  systems and is denoted as $S(\mathbf{p};\mathbf{W}_{S})$, where  $\mathbf{W}_S$ is the parameter set for semantic extraction. Then, the channel encoder, $C(\cdot)$, can be designed separately or jointly with the semantic extractor  and the encoded symbols are generated for channel transmission. The whole encoder process can be expressed as 
\begin{equation}
	\mathbf{s}=C(S(\mathbf{p};\mathbf{W}_{S});\mathbf{W}_{C}),
\end{equation}
where $\mathbf{W}_{C}$ denotes the parameter set for semantic channel encoder.  The transmitted picture can be recovered at the receiver by
\begin{equation}
	\hat{\mathbf{p}}=S^{-1}(C^{-1}(\hat{\mathbf{s}};\mathbf{W}_{C^{-1}});\mathbf{W}_{S^{-1}}),
\end{equation}
where $S^{-1}(\cdot)$ and $C^{-1}(\cdot)$ represent the semantic source decoder and channel decoder, respectively. As indicated in \cite{bao2011towards}, the semantic processing and transmission in the semantic communications are remarkably different from the conventional ones. Meanwhile, the local and shared knowledge in the semantic systems plays a major role. Semantic knowledge can be exploited implicitly or explicitly, as summarized in the following:

1) \textbf{Implicit Semantic Knowledge.} In these designs \cite{zhai2005joint,weng2021semantic,farsad2018deep,xie2020deep}, the local and shared knowledge is implicitly contained in the trainable parameters and the transceivers are usually trained in an end-to-end manner. The impact of physical channels is also learned implicitly. These methods  automatically extract semantic features and cope with the distortion and interference in the physical channels. However,   the trained parameters are difficult to adjust under changing transmit sources or physical environments. 

2) \textbf{Explicit Semantic Knowledge.} In some specific tasks,  the semantic knowledge is shared explicitly. For example,  the semantic network in \cite{yang2021semantic} shares the most important features in the image so that the received image can be classified better than the conventional methods. In \cite{wang2021one}, the photo of the speaker is shared because the appearance of the speaker has no much change during a speech. The explicit shared knowledge  can be adjusted according to the change in the source information, such as replacing the photo for the next speaker.

Apart from the semantic knowledge, the existing methods have not considered adjusting the settings under different channel environments. Therefore, the semantic methods cannot adapt to the  physical channel variation.
\subsection{Link settings in the conventional methods}

 In this  section, we  introduce two key techniques in  wireless communications to cope with changing environments, which can be exploited in semantic system design.
In modern communication systems with HARQ,  the corrupted packets are retransmitted.  IR-HARQ can balance the  requirements of transmission resources and  accuracy and is a popular option. To establish an IR-HARQ system, we need to have a channel encoder and an error detector. If the semantic symbols, $\mathbf{k}$, are protected by a conventional channel encoder, $C(\cdot)$, then the coded symbol can be expressed as
\begin{equation}
	\mathbf{s}=C(\mathbf{k}).
\end{equation}
In \cite{RSMDS}, the  coded symbol vector can be divided into $\mathbf{s}_1$ and $\mathbf{s}_2$ with $\mathbf{s}=[\mathbf{s}_1,\mathbf{s}_2]$ , where $\mathbf{s}_1$ corresponds to coded semantic symbols with high coding rate, and $\mathbf{s}_2$ represent the incremental symbols.    The high code rate symbols, $\mathbf{s}_1$, are transmitted first. Denote $\hat{\mathbf{s}}_1$ as the received symbols corresponding to ${\mathbf{s}}_1$. The recovered semantic symbols can be expressed as,
\begin{equation}
	\hat{\mathbf{k}}=C_1^{-1}(\hat{\mathbf{s}}_1).
\end{equation}
  The conventional CRC error detector is widely used for HARQ systems and extra parity bits are coded from $\mathbf{p}$ and transmitted at the very beginning. With extra parity bits at the receiver, $\mathbf{b}_{\rm CRC}$, ACK information can be generated by
 \begin{equation}
	ACK=Det_{\rm CRC}(\hat{\mathbf{k}},\mathbf{b}_{\rm CRC}), \label{eq10}
\end{equation}
where $Det_{\rm CRC}(\cdot)$ denotes the error detection process.  The feedback signal $ACK=1$ when no error is found; otherwise, $ACK=0$.

The incremental symbols, $\mathbf{s}_2$, need to be transmitted to decrease the code rate if  some errors are founded ($ACK=0$). The received coded symbols are combined together and decoded again, yielding
\begin{equation}
	\hat{\mathbf{k}}=C^{-1}([\hat{\mathbf{s}}_1,\hat{\mathbf{s}}_2]).
\end{equation}
If  the decoded result still has errors, then the retransmission starts, and the above process is repeated. This IR-HARQ method can deal with the varying channels in different time slots.

In addition, diversity on channel conditions at different frequencies of the same time slot can  be exploited. If orthogonal frequency division multiplexing (OFDM) is used, then the overall channel bandwidth can be divided into $L$ parallel flat fading subchannels with different SNRs \cite{abot2012link}.  For OFDM systems, the subchannel gains are different, whereas the noise powers at different subchannels are the same.  Then, the overall SNR can be expressed as 
\begin{equation}
	SNR=\frac{\sum_{l=1}^{L}||h_l\cdot s_l||^2}{\sum_{l=1}^{L}\sigma^2},
\end{equation}
where ${h}_l$ and $s_l$ are the  frequency response and transmit symbols at the $l$-th subchannel, respectively. The $\sigma^2$ is the noise power at the receiver. Once  $\sigma$ and $[h_1,\cdots,h_L]$  are available at the transmitter, the modulation of the $s_l$ can be adaptive to cope with the changing gains of the subchannels. 

Although the combination of the semantic networks and conventional link adaptive methods is naturally considered, the novel mechanism on semantic transmission brings challenges on the design in the technical level. As a deep and inexplicable  network, the performance of the semantic-based transceiver must be guaranteed under varying physical environments.

%
%\subsection{Challenges on Semantic Transceiver}
%Although the combination of the semantic networks and conventional link adaptive methods is naturally considered, the novel mechanism of semantic transmission bring challenges on the design of technical level. As a deep and inexplicable  network, the performance semantic based transceiver must be guaranteed under changing physical environments.
%
% First, the coded semantic symbols are needed to be protected at different extents under subchannels with different noise powers. Then, design a semantic based IR-HARQ framework is also beneficial for different noise powers at different time slots. Finally, the conventional CRC detector need to be replaced by a novel detector to find the  semantic   errors.

\section{Transceiver design for Semantic video conferencing}
In this section, we introduce novel  architectures for semantic video conferencing, which exploit conventional  strategies in wireless communications. We start with a basic network as a semantic source encoder. Then, a novel error detector is proposed to generate an ACK feedback.  The basic network is expanded into the HARQ mode to cope with varying channels at different time slots.  Finally, the CSI for  each subchannel is fed back to the semantic transmitter for adaptive modulation.
\subsection{Basic Semantic Network for Video Conferencing}

Restoring a specific face in an image from few keypoints has been studied in \cite{wang2021one,siarohin2019first}. In these methods, the keypoints contain the changing information of facial expression and manner.  Other information, such as appearance features, does not change during a speech and can be shared to the receiver in advance. Besides,  as presented in \cite{wang2021one},  the keypoints can be compressed and encoded to improve the transmission efficiency. The above methods dramatically reduce the requirement of transmit resource.
However, the existing methods only focus on the framework of source coding and ignore the impact of varying wireless channels. 

\begin{figure}[!h]
	\centering

		\subfloat[ ]{
		\includegraphics[width=5.5in]{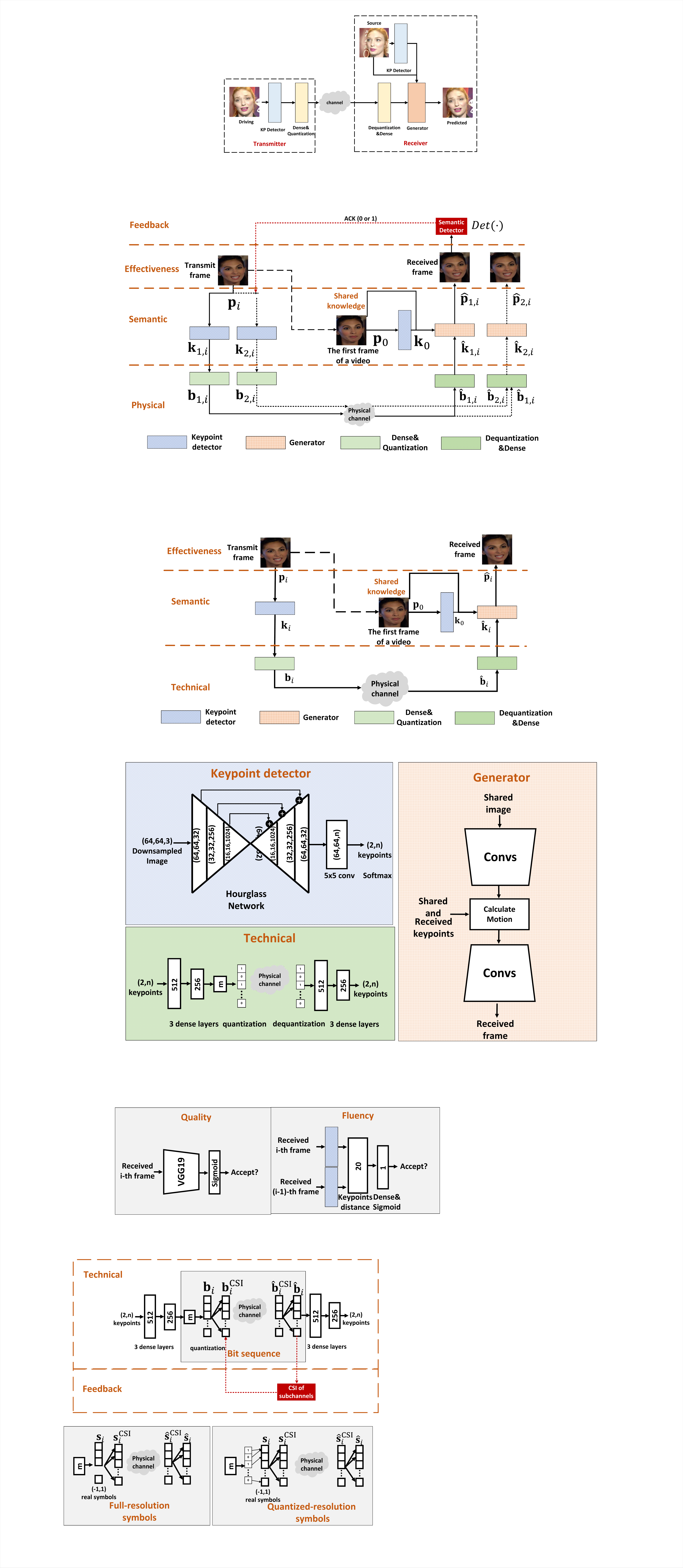}}\\
	\subfloat[ ]{
		\includegraphics[width=5.5in]{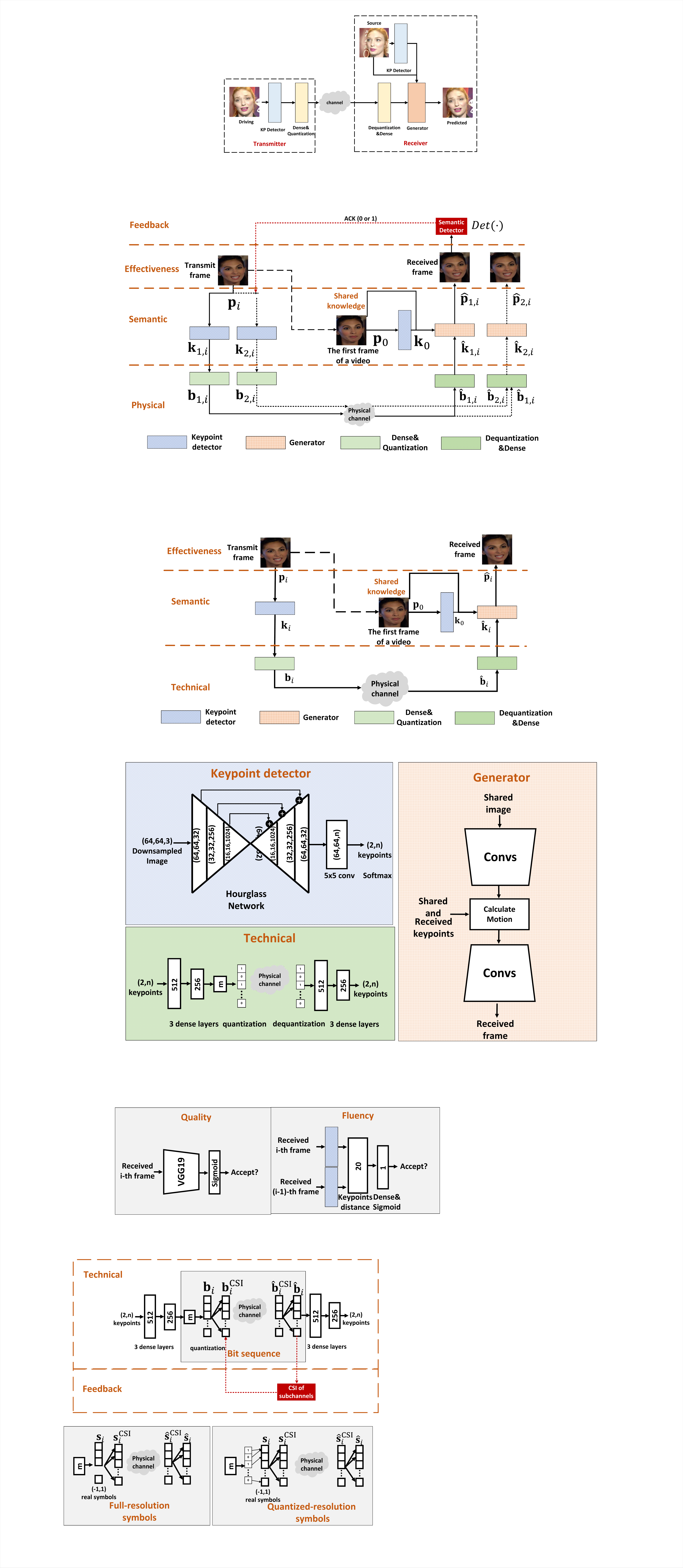}}
	
	% where an .eps filename suffix will be assumed under latex,
	% and a .pdf suffix will be assumed for pdflatex; or what has been declared
	% via \DeclareGraphicsExtensions.
	\caption{(a) Three-level framework of the  semantic video conferencing. (b) Architecture of the keypoint detector and the generator in semantic level, and encoder-decoder in the technical level.  }
	\label{SC_coder}
\end{figure}
 A complete semantic video conferencing framework is shown  in Fig. \ref{SC_coder}, where the simple dense layers are introduced as a channel coding module.  Fig. \ref{SC_coder}(a) shows a semantic video conferencing framework, called SVC. The whole framework consists of three levels similar to \cite{bao2011towards}, including effectiveness, semantic, and technical levels. The effectiveness level   delivers the motion and expression of the speaker. The conventional goal is to minimize the difference of the transmitted and recovered frames. At the semantic level, the photo of the speaker is shared in advance given that the  speaker has no remarkable change during the speech. Usually, the first frame of the video is shared to the receiver for convenience, whereas  a  photo with distinct face is beneficial to generate a good image at the receiver. The keypoint detector extracts the movement of the face in the current frame, and these keypoints are transmitted at the technical level. Based on the received keypoints and the shared photo, the semantic part of the receiver  reconstructs the frame. The networks in the technical level are trained to cope with the distortion and interference from the physical channels. From above description, the SVC has three subnets, including a keypoint detector and a generator in the semantic level, and  an encoder-decoder in the technical level as shown in Fig. \ref{SC_coder}(b). 

The keypoint detector extracts $n$ coordinates of the keypoints, $\mathbf{k}_i\in  \mathbb{R}^{2 \times n}$, from the $i$-th frame, $\mathbf{p}_i\in  \mathbb{R}^{ 256 \times 256 \times 3}$, yielding 
  \begin{equation}
  	\mathbf{k}_i=KD(\mathbf{p}_i;{\mathbf{W} _{\rm KD}}),
  \end{equation}
where $\mathbf{W} _{\rm KD}$ denotes the set of trainable parameters of the keypoint detector.  Specially, the first frame, $\mathbf{p}_0$, with its keypoints $\mathbf{k}_0$, is shared to the receiver. The  keypoint detector consists of convolution neural networks (CNNs) similar to \cite{siarohin2019animating}. The inputted image matrix with sizes (256, 256, 3) is first downsampled to (64, 64, 3) by anti-alias interpolation to   reduce complexity of the keypoint detector. Then, the image is processed by an hourglass network \cite{newell2016stacked} with three blocks.  Each block has a $3 \times 3$ convolution operation with a Relu activation function, a batch normalization, and a $2\times 2$ average pooling. The network has 1024 maximum channels and 32 output channels. After the hourglass network, a $7 \times 7$ convolution converts the output of CNN blocks from (64, 64, 32) into (64, 64, n), thereby dividing the image into $n$ $64\times 64$ grids. The softmax activation is applied to choose grid point with the largest output value. The selected  $n$  grid points are normalized to $[-1, 1]$.

The encoder-decoder  consists of dense  and quantization/dequantization layers. For the encoder, the $n$ keypoints of the $i$-th frame, $\mathbf{k}_i$ (expressed in $n$ coordinates), are considered as $2n$ real numbers and processed by three dense layers, $f_{\rm en}(\cdot)$, with  512, 256, and $m$ neurons, where $m$ is the number of the transmit symbols. The first two layers use Relu activation function, and the last one uses Sigmoid activation function. Then, a two-bit quantization $Q(\cdot)$ is applied to generate $2m$ transmitted bits, $\mathbf{b}_i$. %That means the network learns to transmit the symbols similar to $m$ 16-QAM symbols. 
 The whole process is expressed as
  \begin{equation}
	\mathbf{b}_i=Q(f_{\rm en}(\mathbf{k}_i;{\mathbf{W} _{\rm en}})),
	\end{equation}
where $\mathbf{W} _{\rm en}$ is the set of trainable parameters in the dense layers. 

The dequantizer, $Q^{-1}(\cdot)$, at the receiver in the technical level is the inverse process of $Q(\cdot)$ to recover the $m$ real numbers from the received bits, $\hat{\mathbf{b}}_i$. The three dense layers, $f_{\rm de}(\cdot;\mathbf{W} _{\rm de})$, have  512, 256, $2n$ neurons, where the first two layers use Relu activation function and the last one uses Tanh to restore $n$ keypoints, $\hat{\mathbf{k}}_i$. This process can be expressed as 
  \begin{equation}
	\hat{\mathbf{k}}_i=f_{\rm de}(Q^{-1}(\hat{\mathbf{b}}_i);{\mathbf{W} _{\rm de}}),
\end{equation}
where $\mathbf{W} _{\rm de}$ is the set of trainable parameters in the dense layers.
 The derivative of the quantization layer is replaced by that of the expectation in the backward pass because the gradient  is truncated by the quantization\cite{theis2017lossy}.

The generator reconstructs the current frame from the shared image, $\mathbf{p}_0$, with its keypoints, $\mathbf{k}_0$, and the received keypoints of the $i$-th frame,  $\hat{\mathbf{k}}_i$. This process is denoted as $G(\cdot;{\mathbf{W} _{\rm G}})$, where ${\mathbf{W} _{\rm G}}$ denotes the set of trainable parameters. Therefore, the $i$-th frame can be recovered by   
\begin{equation}
	\hat{\mathbf{p}}_i=G(\mathbf{p}_0,\mathbf{k}_0,\hat{\mathbf{k}}_i;{\mathbf{W} _{\rm G}}),
\end{equation}
where the architecture of  the generator is similar to that  in \cite{siarohin2019first} but without the Jacobian matrix.

The  loss function consists of perceptual loss  \cite{johnson2016perceptual} based on a pretrained CNN, called  VGG-19\cite{simonyan2014very}, a patch-level discriminator loss  \cite{wang2019few}, and an equivariance loss  \cite{siarohin2019first},  denoted as $L_{\rm P}(\cdot)$, $L_{\rm D}(\cdot)$, and $L_{\rm E}(\cdot)$, respectively.  As a result, the overall loss function will be
\begin{equation}
	L(\mathbf{p}_i,\hat{\mathbf{p}}_i)=L_{\rm P}(\mathbf{p}_i,\hat{\mathbf{p}}_i)+L_{\rm D}(\mathbf{p}_i,\hat{\mathbf{p}}_i)+L_{\rm E}(\mathbf{p}_i,\hat{\mathbf{p}}_i).
\end{equation}
Because the trainable parameters in the technical level are much fewer than other parts but  still important, we add a mean-squared-error (MSE) loss function to train the technical level, yielding

\begin{equation}
	L_{\rm MSE}=\frac{\|\mathbf{k}_i-\hat{\mathbf{k}}_i\|^2}{2n}.
\end{equation}
 
 The training processes are divided into three steps. At the beginning, the technical level is ignored, and the parameters in the keypoint detector and the generator are trained by $	L(\mathbf{p}_i,\hat{\mathbf{p}}_i)$, yielding 

\begin{equation}
(\hat{\mathbf{W}}_{\rm KD},\hat{\mathbf{W}}_{\rm G})=\mathop{\arg\min}\limits_{{{\mathbf{W}}_{\rm KD},{\mathbf{W}}_{\rm G}}}L\left(\mathbf{p}_i, G(\mathbf{p}_0,\mathbf{k}_0,KD(\mathbf{p}_i;{\mathbf{W}}_{\rm KD});{\mathbf{W}}_{\rm G})\right).
\end{equation}
Then, the parameters in the technical level are trained by  $L_{\rm MSE}$ to restore the $\mathbf{k}_i$ under the impact of the physical channel distortion, 
\begin{equation}
(\hat{\mathbf{W}}_{\rm en},\hat{\mathbf{W}}_{\rm de})=\mathop{\arg\min}\limits_{{{\mathbf{W}}_{\rm en},{\mathbf{W}}_{\rm de}}}L_{\rm MSE}\left(\mathbf{k}_i, f_{\rm de}(Q^{-1}(Q(f_{\rm en}(KD(\mathbf{p}_i;\hat{\mathbf{W}}_{\rm KD});{\mathbf{W}}_{\rm en})));{\mathbf{W}}_{\rm de})\right).
\end{equation}
Finally, all  trainable parameters of the  SVC are fine-tuned in the end-to-end manner as 
\begin{equation}
(\hat{\mathbf{W}}_{\rm KD},\hat{\mathbf{W}}_{\rm en},\hat{\mathbf{W}}_{\rm de},\hat{\mathbf{W}}_{\rm G})=\mathop{\arg\min}\limits_{{{\mathbf{W}}_{\rm KD},{\mathbf{W}}_{\rm en},{\mathbf{W}}_{\rm de},{\mathbf{W}}_{\rm G}}}L\left(\mathbf{p}_i,\hat{\mathbf{p}_i} \right).
\end{equation}

The proposed SVC is  a  combination of the video synthesis  and theoretic three-level semantic transmission  in wireless communications.  This basic framework is established and trained to study the impact of replacing video transmission with semantic keypoint transmission.  The performance of the  semantic transmission can be  improved further by introducing the ACK feedback in wireless networks, as shown in from the following section.

\subsection{Semantic HARQ with ACK Feedback for Video Conferencing}
 HARQ can cope with time-varying channels in wireless communications. Retransmission and transmitting  incremental symbols are flexible under changing channels with the ACK feedback. Thus, we develop a novel semantic video conferencing framework with HARQ, called SVC-HARQ, to improve semantic transmission.
 
 \begin{figure}[!h]
 	\centering
 		\subfloat[ ]{
 	\includegraphics[width=6in]{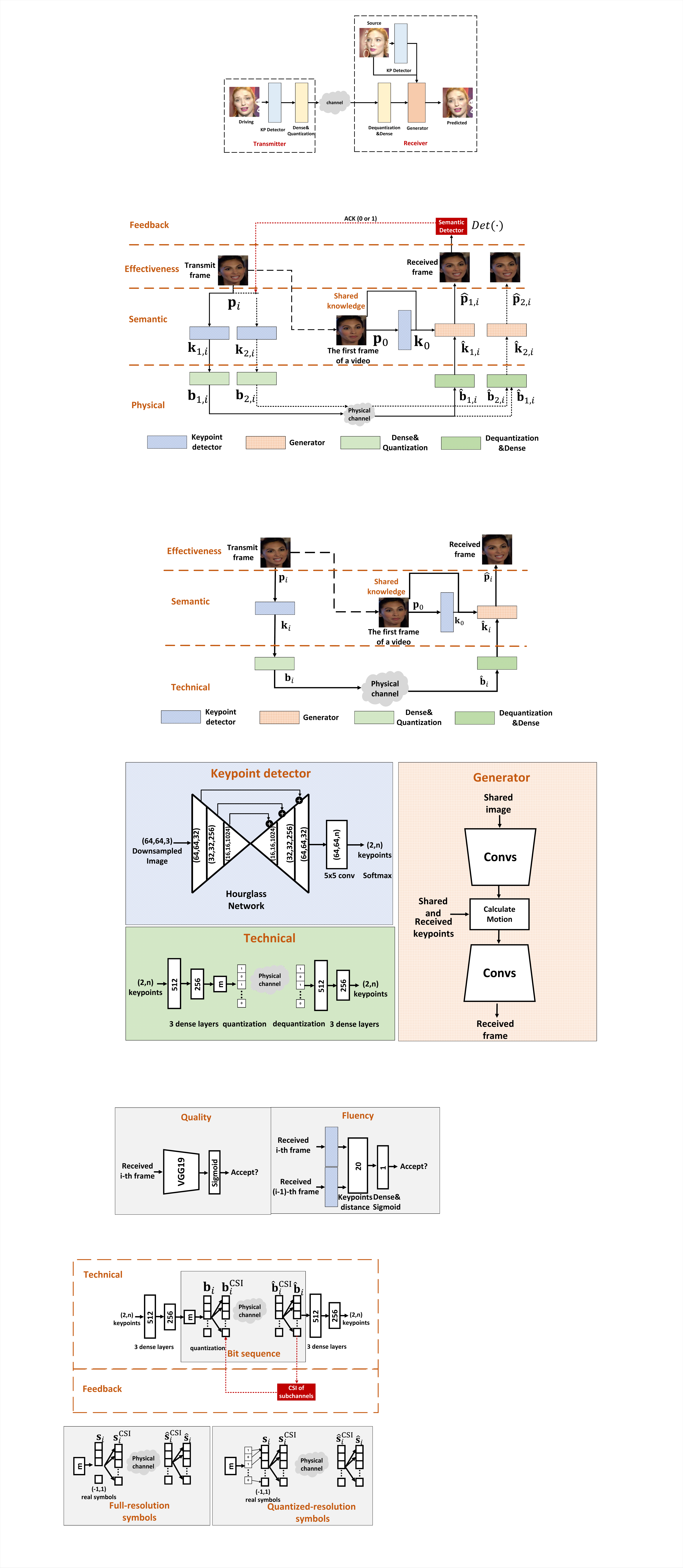}}\\
  		\subfloat[ ]{
 	\includegraphics[width=6in]{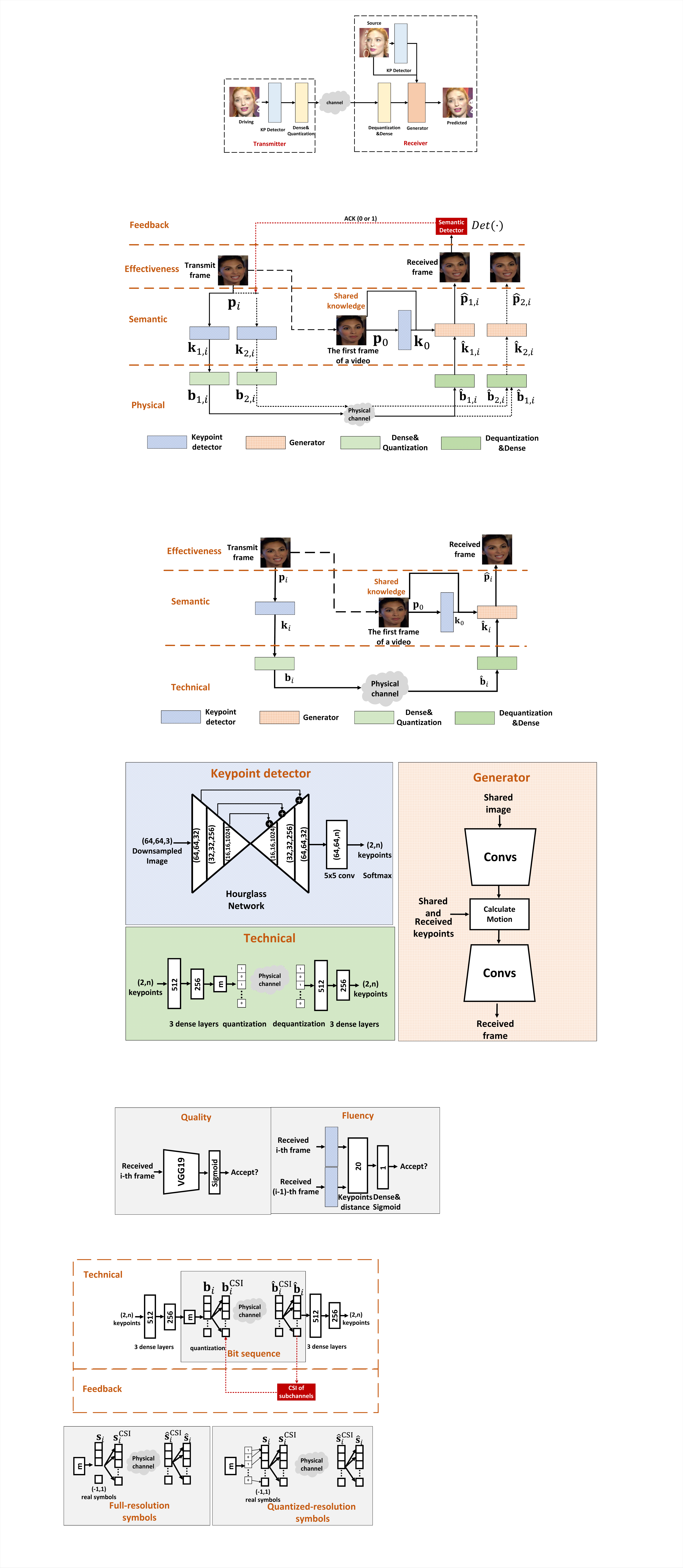}}

 	% where an .eps filename suffix will be assumed under latex,
 	% and a .pdf suffix will be assumed for pdflatex; or what has been declared
 	% via \DeclareGraphicsExtensions.
 	\caption{(a) Structure of SVC-HARQ with ACK feedback. (b) Two potential methods for semantic error detection $Det(\cdot)$.  }
 	\label{SVC-HARQ}
 \end{figure}
 
 As shown in Fig. \ref{SVC-HARQ}, the receiver feeds an ACK signal back to the transmitter after the first transmission. The first transmission is the same as in Fig. \ref{SC_coder} and the trained parameters can be used directly. The first transmitted bit vector, $\mathbf{b}_{1,i}$, can be expressed as
   \begin{equation}
 	\mathbf{b}_{1,i}=Q(f_{\rm en}(KD(\mathbf{p}_i;{\mathbf{W} _{\rm 1,KD}});{\mathbf{W} _{\rm 1,en}})),
 \end{equation}
  where  $\mathbf{W} _{\rm 1,KD}$  is the set of trainable parameters in the keypoint detector, and  $\mathbf{W} _{\rm 1,en}$ is the set of trainable parameters in the encoder.   
  Then, the recovered frame is
\begin{equation}
	\hat{\mathbf{p}}_{1,i}=G(\mathbf{p}_0,\mathbf{k}_0,f_{\rm de}(Q^{-1}(\hat{\mathbf{b}}_{1,i});{\mathbf{W} _{\rm 1,de}});{\mathbf{W} _{\rm 1,G}}),
\end{equation}
where $\mathbf{W} _{\rm 1,G} $ is the set of parameters in the generator of the first transmission, and $\hat{\mathbf{b}}_{1,i}$ represents the received bit sequence at the first transmission. Then, the  reconstructed frame, $\hat{\mathbf{p}}_{1,i}$, is evaluated by a semantic detector. If the detector finds that $\hat{\mathbf{p}}_{1,i}$ is unacceptable, then $ACK$=0  is fed back to the transmitter, and an incremental transmission is triggered. 

The incremental bit sequence is transmitted to correct the errors. Different from the first transmission,  the incremental transmission only concentrates on the fallible keypoints under wicked channel conditions.  Thus, the incremental transmission also needs to be trained and  has  different trainable parameters, namely, $\mathbf{W} _{\rm 2,KD}$, $\mathbf{W} _{\rm 2,en}$, and $\mathbf{W} _{\rm 2,G}$, for the keypoint detector, decoder, and generator, respectively.    The incremental transmitted bit sequence is
\begin{equation}
\mathbf{b}_{2,i}=Q(f_{\rm en}(KD(\mathbf{p}_i;{\mathbf{W} _{\rm 2,KD}});{\mathbf{W} _{\rm 2,en}})),
\end{equation}
 and the recovered frame is
\begin{equation}
\hat{\mathbf{p}}_{2,i}=G(\mathbf{p}_0,\mathbf{k}_0,f_{\rm de}(Q^{-1}([\hat{\mathbf{b}}_{1,i},\hat{\mathbf{b}}_{2,i}]);{\mathbf{W} _{\rm 2,de}});{\mathbf{W} _{\rm 2,G}}),
\end{equation}
where $[\hat{\mathbf{b}}_{1,i},\hat{\mathbf{b}}_{2,i}]$ is the received symbol vector that includes symbols corresponding to  the first and incremental transmission.

The training process of the first transmission is the same as the SVC in Section III A. Then, all trained parameters are used as the initial values when  training the parameters at the incremental transmission.  Besides, the trained parameters, $\hat{\mathbf{W}_{\rm 1, KD}}$ and $\hat{\mathbf{W}}_{\rm 1,en}$, in the first transmission are fixed, and the process can be written as 
\begin{equation}
\begin{aligned}
&(\hat{\mathbf{W}}_{\rm 2,KD},\hat{\mathbf{W}}_{\rm 2,en},\hat{\mathbf{W}}_{\rm 2,de},\hat{\mathbf{W}}_{\rm 2,G})\\
&=\mathop{\arg\min}\limits_{{{\mathbf{W}}_{\rm 2,KD},{\mathbf{W}}_{\rm 2,en},{\mathbf{W}}_{\rm 2,de},{\mathbf{W}}_{\rm 2,G}}}L\left(\mathbf{p}_i, G(\mathbf{p}_0,\mathbf{k}_0,f_{\rm de}(Q^{-1}([\hat{\mathbf{b}}_{1 ,i},\hat{\mathbf{b}}_{2,i}]);{\mathbf{W} _{\rm 2, de}});{\mathbf{W} _{\rm 2, G}}) \right),
\end{aligned}
\end{equation}
where
	\begin{align}
&\hat{\mathbf{b}}_{2, i}=h(Q(f_{\rm en}(KD(\mathbf{p}_i;{\mathbf{W} _{\rm 2, KD}});{\mathbf{W} _{\rm 2, en}}))),\\
&\hat{\mathbf{b}}_{1,i}=h(Q(f_{\rm en}(KD(\mathbf{p}_i;\hat{\mathbf{W}} _{\rm 1, KD});\hat{\mathbf{W}}_{\rm 1, en}))),
\end{align}
and $h(\cdot)$ indicates that the transmitted bits are with random errors due to channel distortion.

 The above description indicates that the semantic detector is the key module of the SVC-HARQ because the detector directly decides whether the incremental transmission or retransmisstion is required. The conventional error detector, CRC, in the HARQ system is unsuitable for the SVC-HARQ because the difference between some subtle errors in the received frames are acceptable for the conferee. We use an image quality assessment method \cite{zhang2018blind} to evaluate whether or not the received frame is acceptable.                                                                                                                                                                                                                                                                                                                                                                                           This quality assessment network can be obtained by transfer-learning a VGG-19 based classifier as shown in the left of Fig. \ref{SVC-HARQ}(b).   The VGG-19 based quality detector consists of VGG-19  and one dense layer with Sigmoid activation function to output frame quality indicator. The received frame is labeled as 1 for acceptable quality and 0 for unacceptable quality. The loss function is cross-entropy. With a trained detector, $Det_{\rm VGG}(\cdot)$, the ACK feedback can be expressed as
 \begin{equation}
 ACK=\left\{
 \begin{aligned}
1, &~Det_{\rm VGG}(\hat{\mathbf{p}}_{1,i})>0.5,\\
0, &~Det_{\rm VGG}(\hat{\mathbf{p}}_{1,i})\leq 0.5.
 \end{aligned} \right.  \label{ACKcal}
 \end{equation}
 
 In fact, the errors in the received keypoints dp not decrease the image quality directly, but they change the facial expressions.  The generator can reconstruct an acceptable  face image even if the keypoints have some errors because the general appearance is obtained from the shared image. The error keypoints  only change the current expression and lead  the video to be  not fluent. To detect these changes, we propose a novel fluency detector on the right of Fig. \ref{SVC-HARQ}(b).    Thus, the detector needs to distinguish  inappropriate expressions. We use a keypoint detector to capture the keypoints after $\hat{\mathbf{p}}_{1,i}$. Then, we calculate the distance between the keypoints of $\hat{\mathbf{p}}_{1,i}$ and $\hat{\mathbf{p}}_{1,i-1}$. A large distance means that the expression has a sudden change and the transmitted keypoints have some errors. The whole detection process can be expressed as

  \begin{equation}
Det_{\rm KD}(\hat{\mathbf{p}}_{1,i-1},\hat{\mathbf{p}}_{1,i})=f_{\rm Det}((KD(\hat{\mathbf{p}}_{1,i-1})-KD(\hat{\mathbf{p}}_{1,i}))^2;\mathbf{W}_{\rm Det}),
 \end{equation}
 where $f_{\rm Det}(\cdot)$ is a dense layer with one neuron output and Sigmoid activation function. This detector is trained with cross-entropy as the loss function by the reconstructed frames collected by the SVC under different channels. The average keypoint distance (AKD) calculated by a pretrained facial landmark detector \cite{bulat2017far} is used for labeling, where the output of the detector is labeled  with 1  if its AKD is smaller than five and 0 otherwise.  The loss function is cross-entropy. After training, the fluency detector is also used similar to Eq. (\ref{ACKcal}). 

The proposed quality and fluency detectors are  compared  for HARQ systems under semantic-based video conferencing.    Overall, an extra incremental transmission process can make the SVC more adaptive under changing channels if the semantic detector is  effective. Moreover, the retransmission is started if the incremental symbols cannot correct the errors to reach the criterion of the detector.

\subsection{Adaptive Encoding with CSI Feedback}
\begin{figure}[!h]
	\centering

	\subfloat[ ]{
		\includegraphics[width=6in]{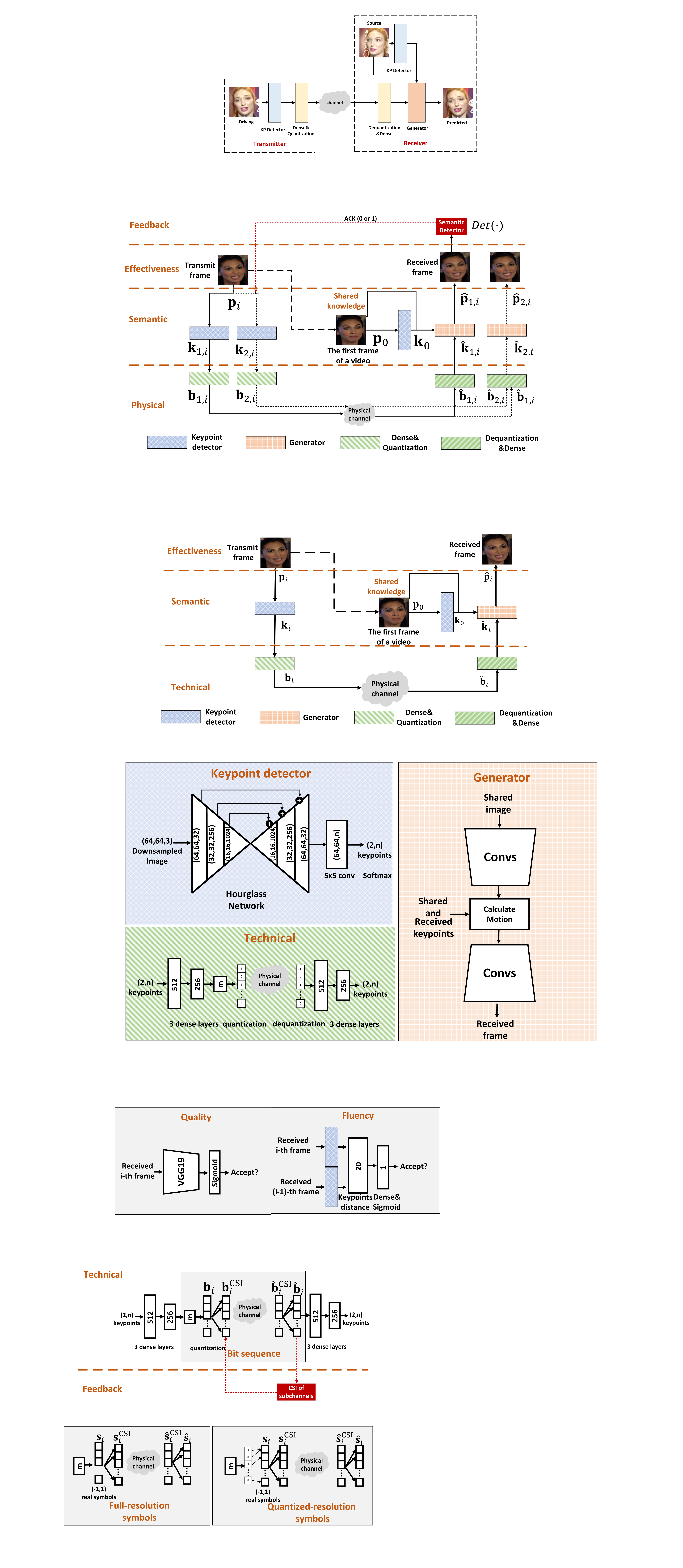}}\\
	\subfloat[ ]{
		\includegraphics[width=6in]{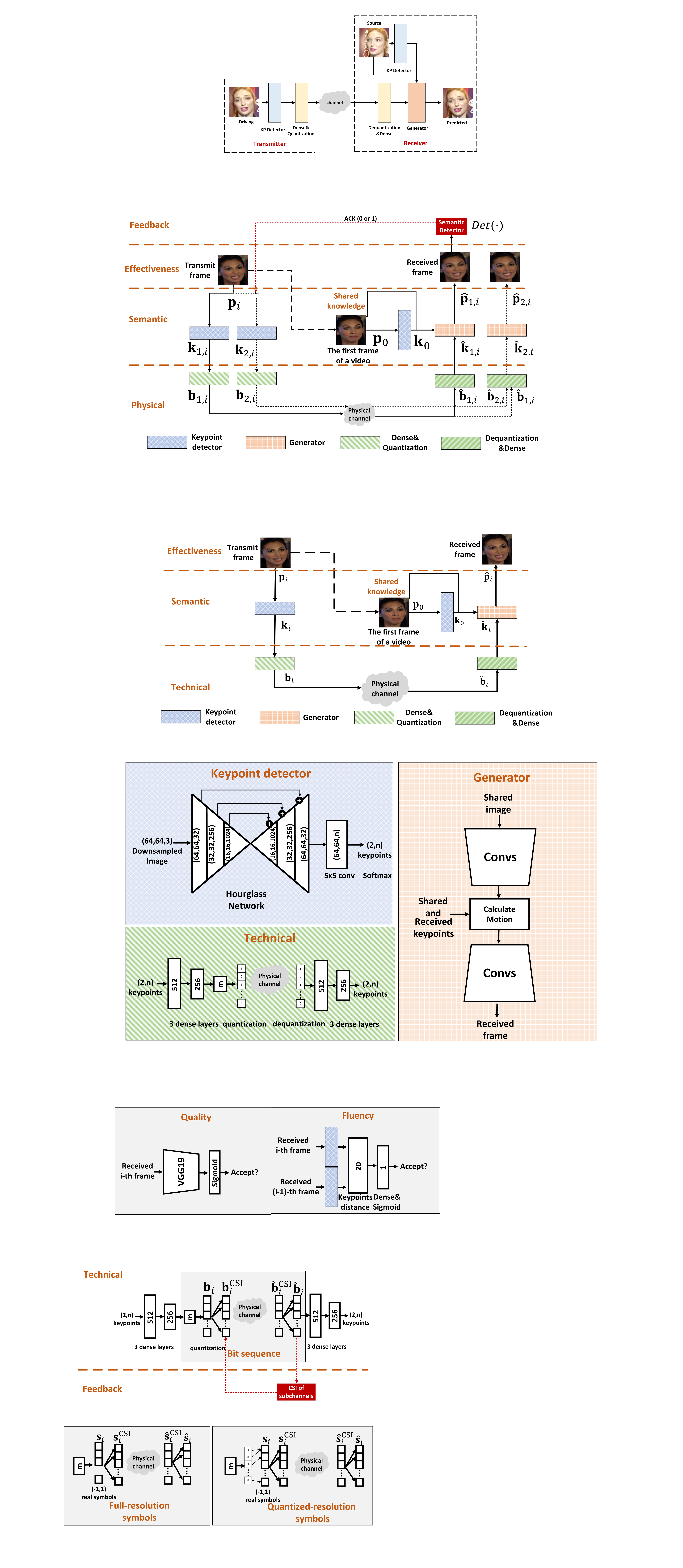}}
	
	% where an .eps filename suffix will be assumed under latex,
	% and a .pdf suffix will be assumed for pdflatex; or what has been declared
	% via \DeclareGraphicsExtensions.
	\caption{(a) Structure of SVC with CSI feedback.  (b) Two different modulation methods for SVC-CSI. }
	\label{SVC-CSI}
\end{figure}
The above SVD methods exploit no CSI further. However, the noise power of the subchannels can be obtained by the receiver. For example, the frequency selective channels can be divided into different subchannels with different SNRs. We assume that the CSI of all  subchannels  is estimated by the receiver and shared to the transmitter. These channel conditions are exploited by the encoder-decoder at the technical level, which helps to protect the most important keypoints. The accurate CSI of each subchannel cannot be obtained in practice, and the feedback of the entire CSI values also requires resources. Thus, the receiver sorts the subchannels by their  channel conditions and feeds this sequence back to  the transmitter. This method simplifies the design of encoder-decoder in the technical level and reduces the feedback cost.

Compared with the original  SVC, that with CSI feedback (SVC-CSI) only needs to add a sort module $SN(\cdot)$ as 
shown in  Fig. \ref{SVC-CSI}(a).The output of the sort module is denoted as $\mathbf{b}_{i}^{\rm CSI}=SN(\mathbf{b}_{i})$, where its elements, representing subchannel gains, are in decreasing order. Then, $\mathbf{b}_{i}^{ \rm CSI}$ is sent to the receiver.  At the receiver,  received $\hat{\mathbf{b}}_{i}^{ \rm CSI}$ is restored by $SN^{-1}(\cdot)$.
%, yielding
%   \begin{equation}
%\hat{\mathbf{b}}_{i}=SN^{-1}(\hat{\mathbf{b}}_{i}^{ \rm CSI})=SN^{-1}([\cdots,\hat{\mathbf{b}}_{i,1},\cdots])=[\hat{\mathbf{b}}_{i,1},\cdots,\hat{\mathbf{b}}_{i,L}].
%\end{equation}
Because the other parts are the same as the SVC and the sort module of the SVC-CSI has no impact on the gradient, the SVC-CSI has the same training strategy as the SVC.

The above methods encode the keypoints into bit sequence, which can be easily applied into the conventional wireless communication systems, such as the OFDM system with quadrature amplitude modulation (QAM). Furthermore, joint design with modulation module to encode the keypoints into constellation points directly   can further improve the performance.   Thus, we also investigate the benefit of CSI feedback on the encoding keypoints  into constellation points. In the left structure of  Fig. \ref{SVC-CSI}(b), we  directly replace the quantization in Fig. \ref{SVC-CSI}(a)  with a dense layer and Tanh activation function. Its output has $m$ real symbols, which denote $m/2$  constellation points. These points are also rearranged according to CSI feedback. This constellation method is called full-resolution because the learned  constellation points can appear anywhere in the constellation.

However,  the full-resolution constellation is extremely complex for practical systems due to the finite precision. Thus, the  constellation points need to be limited. We combine two bits into a real symbol similar to 16QAM. Meanwhile, each 2-bit vector is coded by the shared two trainable parameters, $\alpha$ and $\beta$, yielding

   \begin{equation}
s_{i,j}=\alpha b_{i,2j-1}+\beta b_{i,2j}, j=1, \cdots, m,
\end{equation}
where the $2m$ bits in $\mathbf{b}$ are first modulated into the real symbols, $s_{i,j}$, which only have four possible values, i. e.,  the constellation points only appear in 16 locations.
The $m$-symbol vector $\mathbf{s}_i$ is also divided into $L$ subchannels and has been multiplied to different transmit powers, $\bm \rho=[\rho_1,\cdots,\rho_L] $. 
%This process can be expressed as
%   \begin{equation}
%\mathbf{s}_i=[\rho_1 \mathbf{s}_{i,1}, \cdots, \rho_L \mathbf{s}_{i,L}].
%\end{equation}
Then, $\mathbf{s}_i$ is rearranged to  $\mathbf{s}_{i}^{\rm CSI}$   according to CSI feedback and sent to the receiver. The training process of these two methods with constellation points is still the same as that of  SVC. Specially, this method is called quantized-resolution and only introduces $L+2$ parameters $\alpha$, $\beta$, and  $\bm \rho$.

In general, some  bits/symbols  always transmit under better channel conditions than the others with CSI feedback. Thus, the networks can learn to transmit  important keypoints at the  subchannels with high SNRs.

\section{Numerical Results}
	In this section,  we  present the numerical results of different  frameworks  and   discuss  the pros and cons of the semantic-based video conferencing. We  also compare their bit consumption (required number of bits) with competing ones.

\subsection{Configurations of the simulation system}
\textbf{Training settings.} The VoxCeleb dataset \cite{nagrani2017voxceleb} has  considerable face videos of speakers.  These videos are pre-processed into the size of 256 $\times$ 256, and the videos without a distinct face are removed. Meanwhile, all the videos have only one speaker.   After pre-processing, the training dataset has 2000 videos  and the testing dataset has 100 videos, which have a total of about 500 different  speakers.  All  networks are trained with Adam optimizer \cite{kingma2014adam}, and their initial learning rate is 0.0002. 

\textbf{Baseline.} H264 is  widely used as a commercial standard and usually occupies one-tenth bandwidth of the original video. The constant rate factor of the H264 code can be adjusted to generate different qualities and sizes of videos. Meanwhile, RS code is commonly applied in storage and channel coding in the wicked environment, such as deep space communications. In the following, RS$(n, k)$ means to encode $k$ information symbols into  $n$  symbols. The redundancy $(n-k)$ symbols can correct $(n-k)/2$ error symbols. In this study,  IR-HARQ strategy\cite{MDS} encodes 64 information symbols into 255 symbols and transmit 127 symbols initially. The remaining 128 symbols are transmitted as the incremental redundancy, whereas the CRC detects errors after the first transmission.

\textbf{Metrics.}  Three metrics are used to  evaluate the results:

1) \textbf{Average keypoint distance (AKD).} We use a pretrained facial landmark detector \cite{bulat2017far} to evaluate the errors in our transmission. This pretrained detector extracts keypoints from the received  and transmitted frame, and their average distance is computed. The AKD metric represents the motion and the changing expression of the face.

2) \textbf{Structural similarity index measure (SSIM).} SSIM evaluates the structural similarity among patches of the input images \cite{hore2010image}. Therefore, SSIM, which is more robust than PSNR, is widely used as the metric of images.

3) \textbf{Perceptual loss (Ploss).} Perceptual loss is commonly used as a regularization method when training a network in the computer vision. Through calculating the sum of MSEs between the estimated and the true image at the different layers of a pretrained network, such as VGG,  the similarity of the features represented by the perceptual loss. Here, we choose the perceptual loss metric proposed in \cite{zhang2018unreasonable}.

\subsection{Performance of semantic coding}

%	\begin{figure}[!h]
%		\centering
%		
%
%			\includegraphics[width=5.5in]{figure//baseline}
%
%		
%		% where an .eps filename suffix will be assumed under latex,
%		% and a .pdf suffix will be assumed for pdflatex; or what has been declared
%		% via \DeclareGraphicsExtensions.
%		\caption{The performance of the SVC and H264. The H264 has different qualities under different settings of average bits per pixel (Bpp). }
%		\label{Metric1}
%	\end{figure}

In TABLE \ref{Metric1}, the SVC only transmits 160 bits per frame and the conventional H264 coding always has better SSIM performance than the SVC. However,  the H264 requires more bits per pixel. Then, the two methods are compared in the metrics of Ploss and AKD, where the facial features are more important  than the structural similarity. The SVC has a tremendous superiority in the bit consumption.  The required bits per pixel of the H264 are about six times those of the SVC when   their Ploss   performances are similar and more than four times those of the SVC when their AKD performances are similar. In general, these metrics only  represent different perspectives in the evaluation of semantic transmissions and the transmitted results should also be acceptable to  humans. In the following section, we analyze the semantic errors through examples. The H264 with 0.0157 Bpp (bold in the table) is selected as the baseline and  called H264 for convenience. Thus, the Ploss performance of the H264 and the SVC is the same without the impact of the channels. The AKD and SSIM metrics of the H264 are   better than those of the SVC.  

  	\begin{table}[!h]
	\centering	
	\caption{The performance of the SVC and H264. }
	\begin{tabular}{>{\sf }c|c|lllllll}    %
		\toprule
		\multirow{4}{*}{H264}	&	Bpp& 0.0127 & 0.0136&	\textbf{0.0157}&0.0185&0.0221&0.0261&0.0316\\ \cline{2-9}
		&	SSIM $\uparrow$	&0.759	&0.780 &		0.803&	0.826&	0.847&	0.865 &	0.882 \\ \cline{2-9}
		
		&	Ploss $\downarrow$	&0.227 &	0.201&	\textbf{0.187}&	0.159&	0.136	&0.116&	0.094\\ \cline{2-9}
		&	AKD $\downarrow$&3.23 &	2.85&		2.50&	2.27&	2.08&	2.01&	1.89\\ 
		
		\bottomrule
		\multicolumn{9}{c}{}

	\end{tabular}
	
	\begin{tabular}{>{\sf }c|ll}   
		\toprule
		\multirow{4}{*}{SVC} &Bpp&\textbf{0.0024}   \\ \cline{2-3}
		&SSIM$\uparrow$&0.671\\ \cline{2-3}
		&Ploss$\downarrow$ &\textbf{0.186}\\ \cline{2-3}
		&AKD$\downarrow$ &3.12\\ 
		
		\bottomrule
	\end{tabular}
	\label{Metric1}
\end{table}

\begin{figure}[!h]
	\centering
	
\subfloat[ ]{ 
		\includegraphics[width=2in]{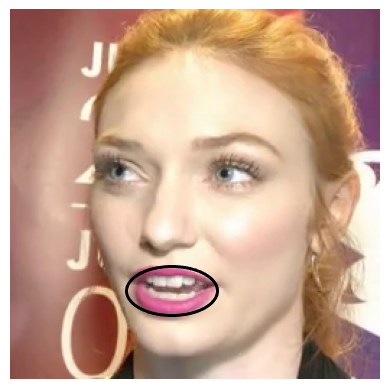}}
	\subfloat[ ]{ 	\includegraphics[width=2in]{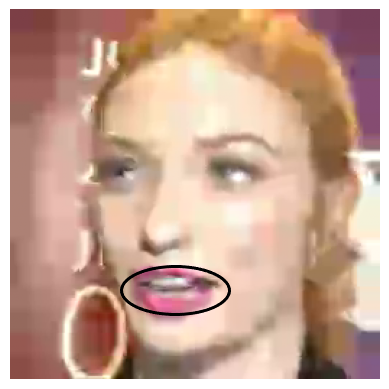}}
	\subfloat[ ]{ 	\includegraphics[width=2in]{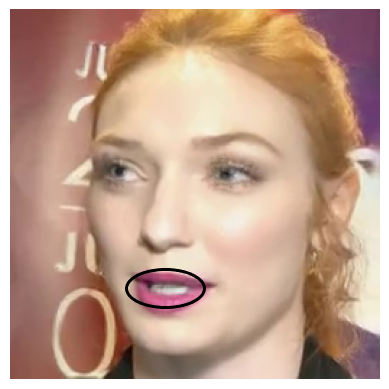}
	}

	% where an .eps filename suffix will be assumed under latex,
	% and a .pdf suffix will be assumed for pdflatex; or what has been declared
	% via \DeclareGraphicsExtensions.
	\caption{Different content loss after the conventional H264 and the SVC coding. Their Ploss performances are similar. (a) Original frame.  (b) Frame coded by the H264. (Ploss=0.187) (c) Frame coded by the SVC. (Ploss=0.186)  }
	\label{SVC1}
\end{figure}

Although the H264 has the same Ploss performance as the SVC, the content loss  is different, as shown in the examples in  Fig. \ref{SVC1}. The H264 loses the pixel information in all the areas of this frame; thus, the  frame shows lower resolution than the original one. The lost information of the SVC usually cannot be distinguished as an independent image. Only the detailed expressions in the frame coded by SVC, such as the mouth in the circles, are different from the original because the semantic information is ignored when coding. This phenomenon is demonstrated in three metrics in Table \ref{Metric1}. Considered as a lower resolution image of the original frame, the structural information of the H264 frame is still reserved and the locations of the detected keypoints   unchanged. Thus, the H264 has better SSIM and AKD metrics than the SVC.  However, the quality of the SVC frame seems higher than the H264 to the human if the detailed expressions lead no ambiguities. Considering that the SVC only requires 1/6 of the bits of the H264, the semantic transmission is a better option for video conferencing, especially some conferees are using mobile phones in the crowd. 
	\begin{figure}[!h]
	\centering	
	
	%	\subfloat[ ]{ 	\includegraphics[width=2.7in]{figure//H264RS1}}
		\subfloat[ ]{ 	\includegraphics[width=3in]{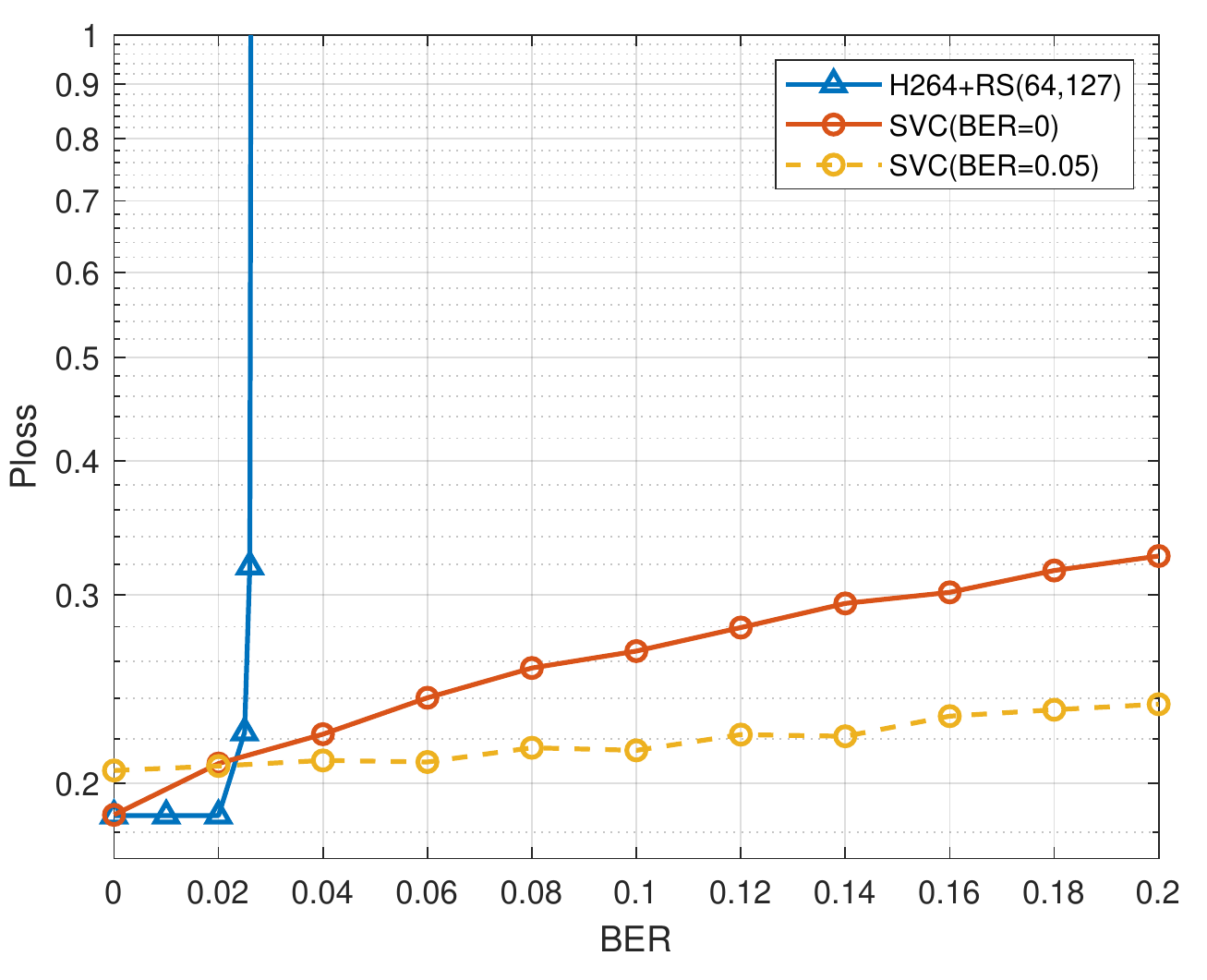}}
		\subfloat[ ]{ 	\includegraphics[width=3in]{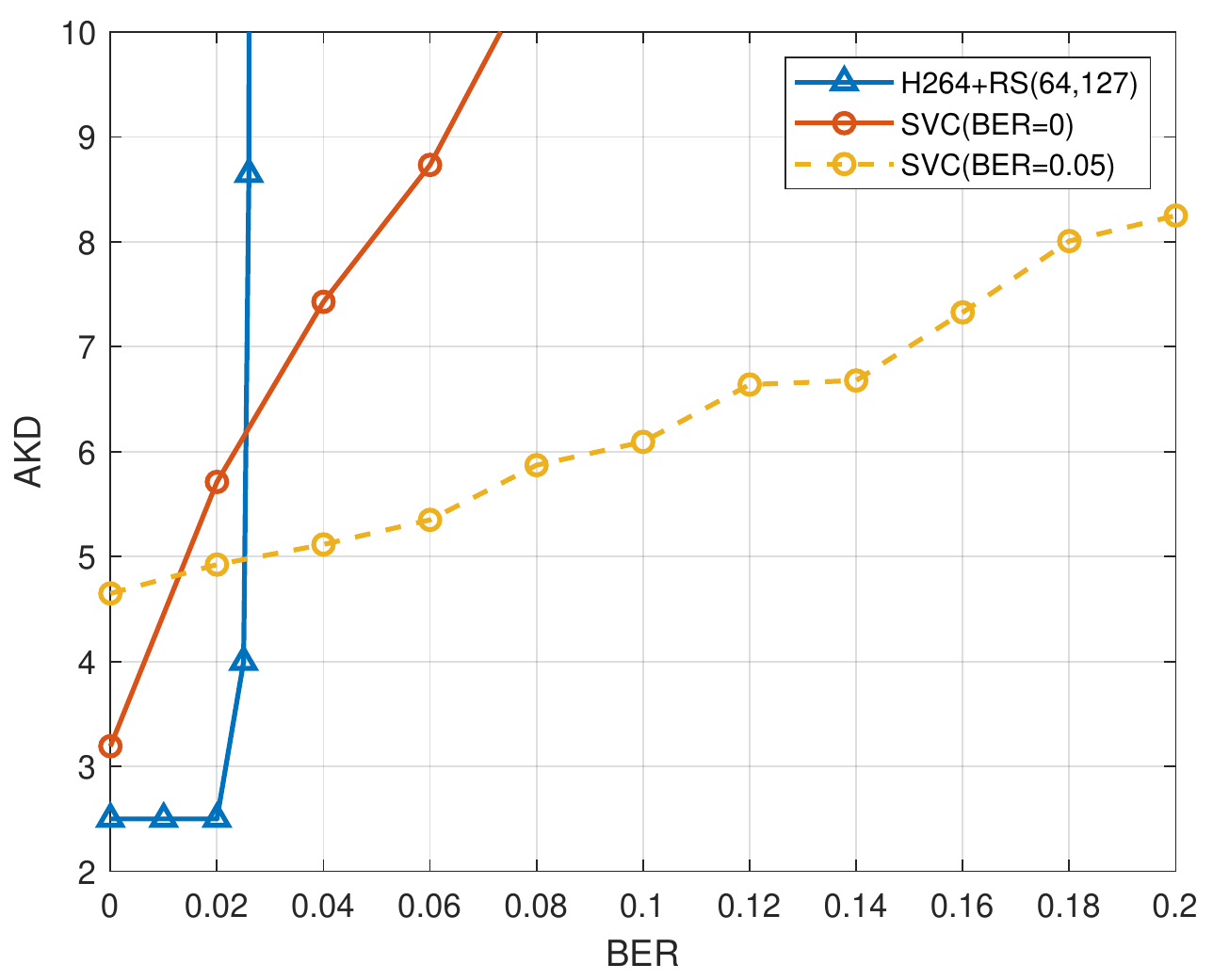}}
	
	% where an .eps filename suffix will be assumed under latex,
	% and a .pdf suffix will be assumed for pdflatex; or what has been declared
	% via \DeclareGraphicsExtensions.
	\caption{Performance of the SVC trained under different BERs. The competing method is encoded by the H264 and the RS channel coding. }
	\label{SVC2}
\end{figure}

As shown in Fig. \ref{SVC2}, the two methods have different sensitivities to the change of BER. The conventional H264+RS can perfectly restore the transmitted information when the errors are fewer than its correction capability. Thus, the performance of the H264+RS is unchanged when BER$=0\sim 0.02$ and decreases sharply when BER $=0.025\sim 0.027$. The SVC methods always have better performance in terms of the three metrics when BER $>0.027$ because the semantic transmission can still repair the semantic errors under high BER. However, the training environment is important for the performance of the SVC. The SVC (BER$=0$) is more suitable for low BER, and its performance becomes worse than the SVC (BER$=0.05$) when BER $>0.02$. That means the two SVC methods allocate different transmit resources for coping with errors implicitly.

In general, the SVC can save transmit resources for a high resolution video conferencing because it only transmits keypoints and has no need to compress pixel information such as H264. Moreover, the SVC has a superiority  under the extremely high BER. However, the training BER affects the performance of the SVC, similar to selecting the code rate of channel coding. Thus, the IR-HARQ frame of the SVC is proposed and tested in the following section.

\subsection{Performance of SVC-HARQ}

 \begin{figure}[!h]
	\centering

	\subfloat[ ]{ 
		\includegraphics[width=1.8in]{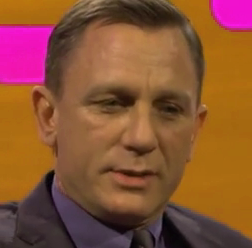}}
	\subfloat[ ]{ 
		\includegraphics[width=1.8in]{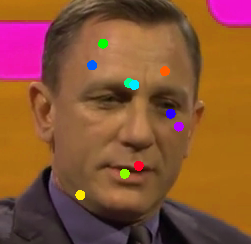}}\\
	\subfloat[ ]{ 
		\includegraphics[width=1.75in]{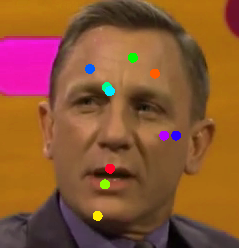}}
	\subfloat[ ]{ 
		\includegraphics[width=1.8in]{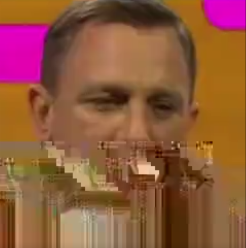}}
	
	% where an .eps filename suffix will be assumed under latex,
	% and a .pdf suffix will be assumed for pdflatex; or what has been declared
	% via \DeclareGraphicsExtensions.
	\caption{ Error examples of the conventional and semantic methods  under BER=0.05.  (a)  Source frame.  (b) Transmit keypoints. (c) Received keypoints and restored frame. The AKD of the semantic method is about 9. (d) Received frame with H264+RS(64, 127). The keypoints  cannot be detected to calculate AKD because the face is blurred. }
	\label{Serr}
\end{figure}
	
 Before discussing the performance of the SVC-HARQ, we first take a look at  the difference between the semantic and bit errors. As shown in Fig. \ref{Serr},  the bit errors in the H264+RS(64, 127) blurry the frame directly and even cause the speaker to become unrecognizable, where  no  semantic errors are found in the same channel condition independently. In fact, these keypoints of the received frame  in the Fig. \ref{Serr}(a) are out of position. Apart from the loss of detailed expressions in Fig. \ref{SVC2}, the transmission errors  lead to the change in the expressions. Thus, an effective error detector should be proposed to guarantee the quality of the SVC-HARQ.

  	\begin{figure}[!h]
 	\centering	
 	
 	\subfloat[ ]{ \includegraphics[width=3in]{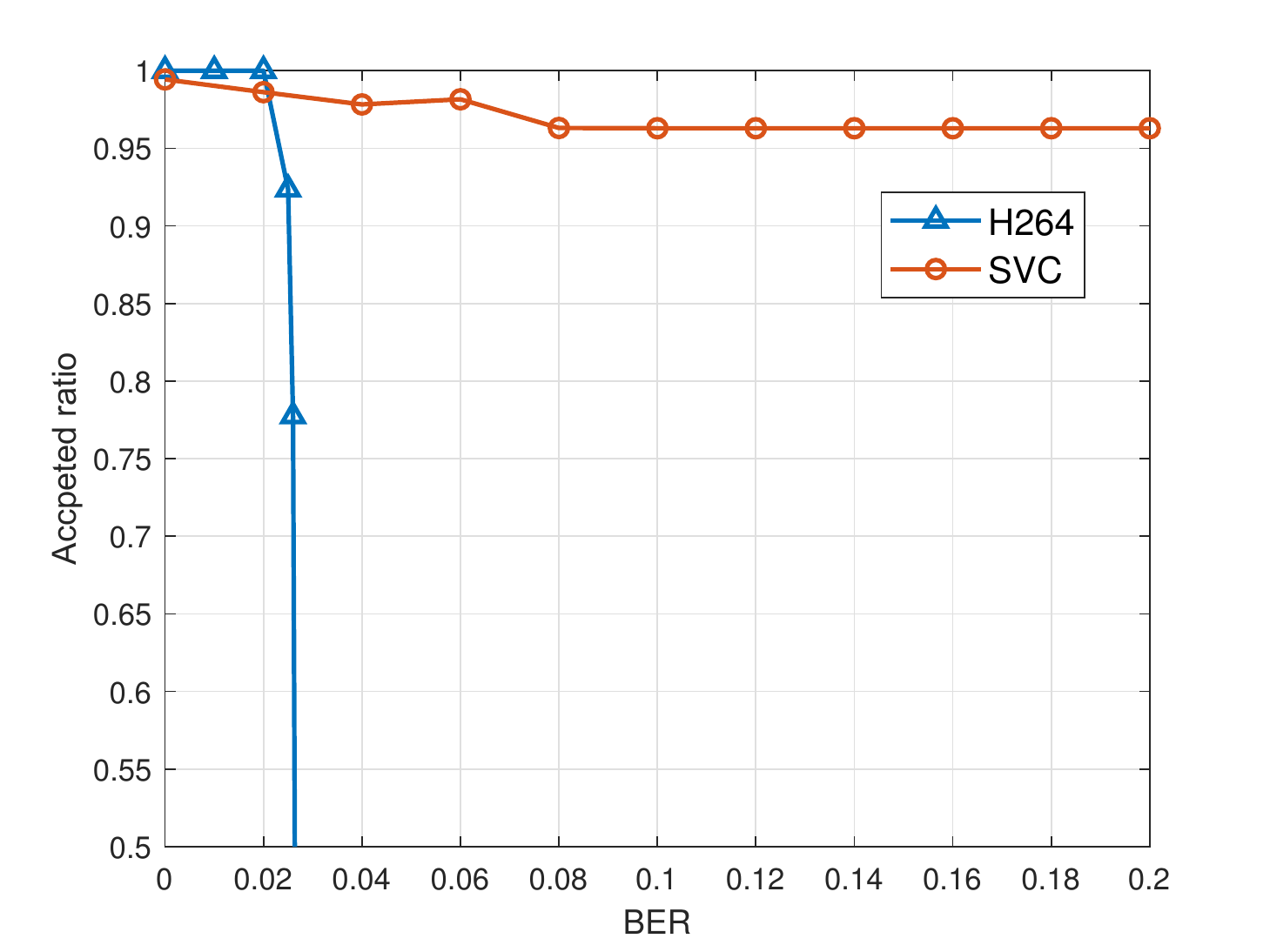}}
 	\subfloat[ ]{ 
 	\includegraphics[width=2.9in]{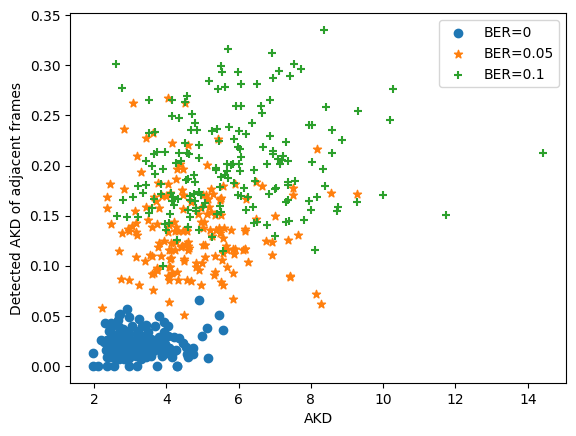}}
 	
 	% where an .eps filename suffix will be assumed under latex,
 	% and a .pdf suffix will be assumed for pdflatex; or what has been declared
 	% via \DeclareGraphicsExtensions.
 	\caption{(a) Accepted ratio of the received frames  using VGG-based detector under different BERs. (b) Detected AKD of the adjacent frames and their AKD performances under different BERs. }
 	\label{Detacc}
 \end{figure}

The errors in the SVC are difficult to find independently; thus, the VGG-based quality detector always achieves higher accepted ratio than 96\%. Therefore, the SVC can protect the visual quality even under the wicked environment	because the main appearance features are shared in advance. Thus, the VGG-based quality detector is insufficiently effective as a semantic detector. Apart from the quality of the received frame, the received video should be fluent.    Meanwhile, the performance of the current frame cannot be obtained because the true current frame is unknown to the receiver in practice. Thus, the keypoints are detected again at the receiver by the trained keypoint detector, and the average distances of the detected keypoints between the adjacent frames, called detected AKD, are related to the video fluency. As shown in Fig. \ref{Detacc}(b), the detected AKDs  of the most frames are lower than 0.05 when   no bit error exists and increase with   BER. Compared with the 32-bit CRC code used in the conventional HARQ systems, the fluency detector is helpful to guarantee the quality of the video without any extra parity code.

 	\begin{figure}[!h]
 	\centering	
 		\subfloat[ ]{ 
 	\includegraphics[width=3.08in]{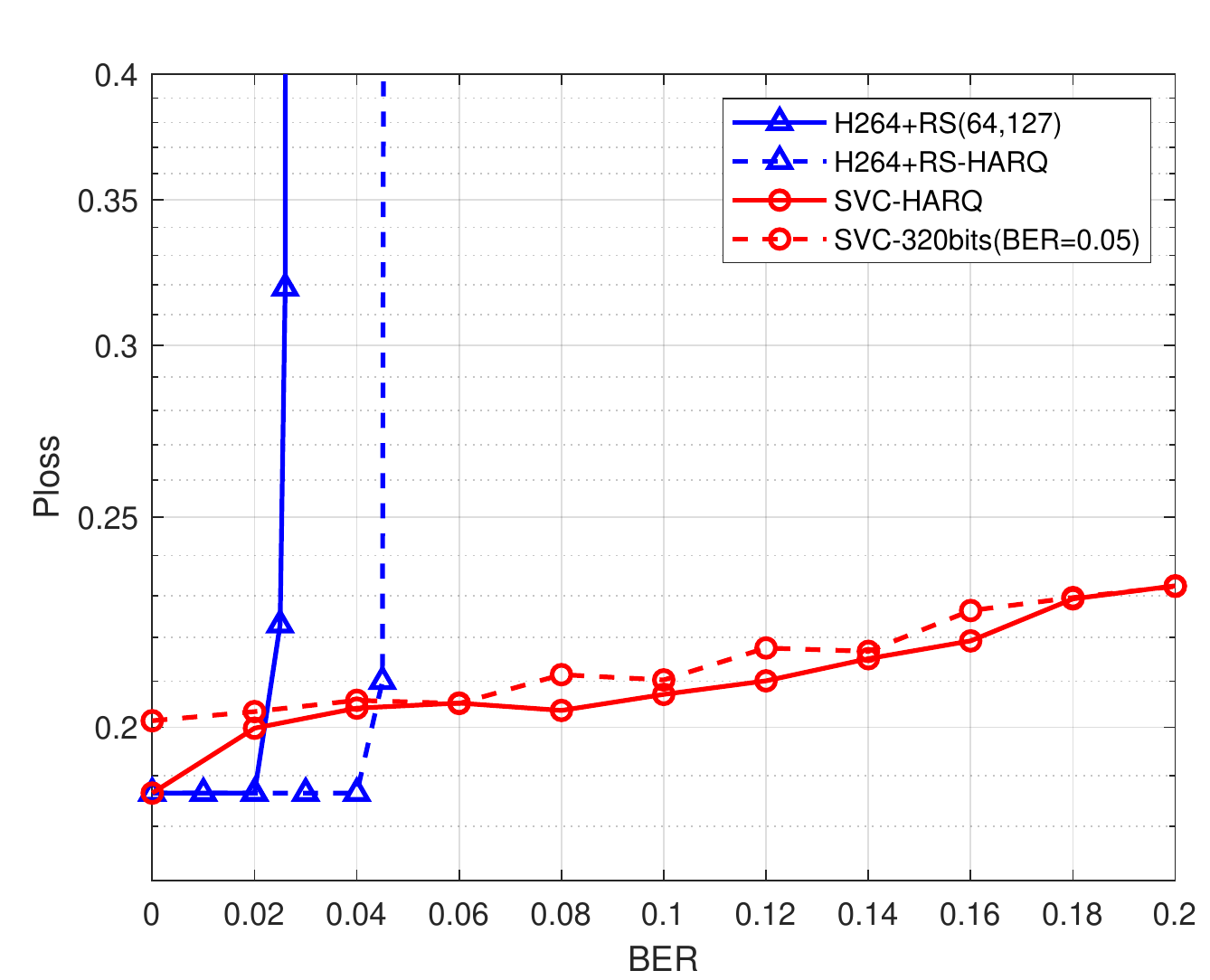}}
 \subfloat[ ]{ 
 	\includegraphics[width=3in]{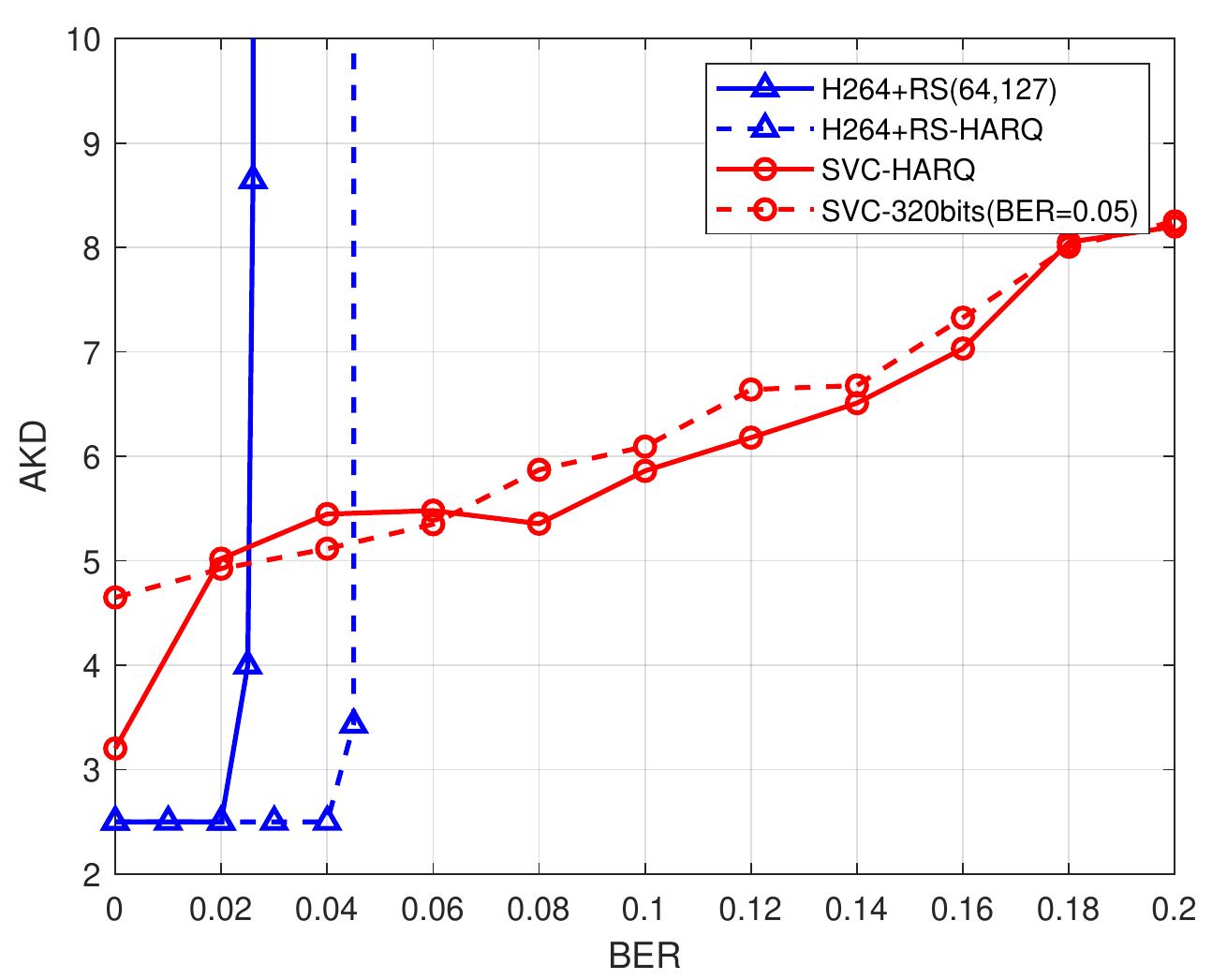}}\\
 	\subfloat[ ]{ 
 	\includegraphics[width=4in]{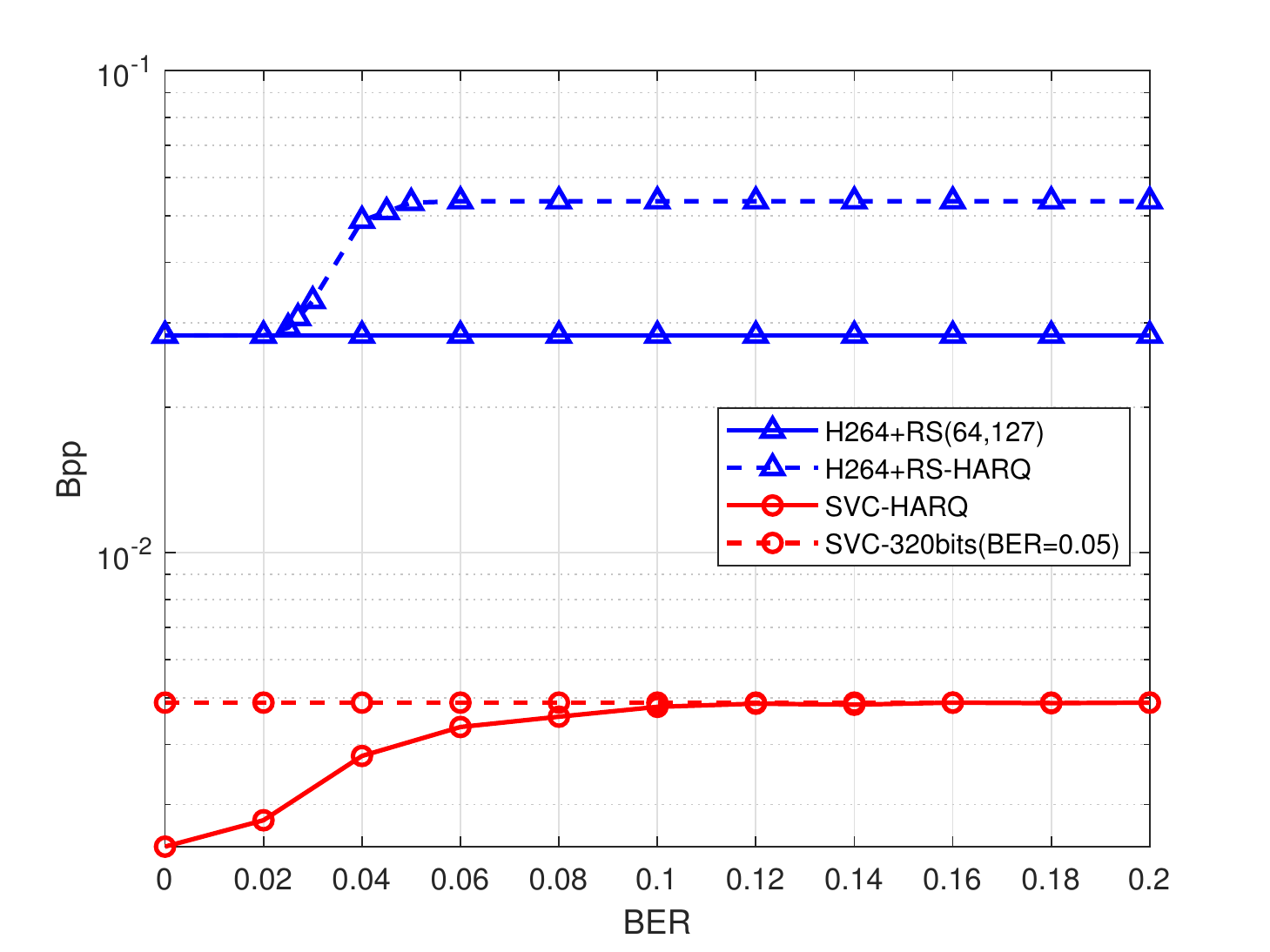}}
 	
 	% where an .eps filename suffix will be assumed under latex,
 	% and a .pdf suffix will be assumed for pdflatex; or what has been declared
 	% via \DeclareGraphicsExtensions.
 	\caption{(a) Ploss performance of SVC-HARQ  and the competing methods. (b)  AKD performance of SVC-HARQ  and the competing methods. (c) Bit consumption of SVC-HARQ  and the competing methods. }
 	\label{SVCHARQ}
 \end{figure}
 
 % Then, it learns 200 incremental bits for the first retransmission which has similar bit limit as SC. The second retransmission contains 500 incremental bits which ensures 99\% word accuracy when BER=0.05. In Fig. \ref{SCHARQ}(a), the SC HARQ is much better than conventional methods when BER $\geq$ 0.1. However, it also cannot achieve 100\% word accuracy when BER=0 and their sentence error rate is higher than the conventional methods when BER=0-0.04.
Finally, the whole SVC-HARQ framework is tested as shown in Fig. \ref{SVCHARQ}. The competing method, H264+RS-HARQ, first transmits 127 symbols per 64 information symbols. Then, the incremental 128 symbols are transmitted if the CRC detector finds errors. Compared with  H264+RS(64, 127), H264+RS-HARQ can correct more errors and maintain its best performance when BER is no larger than 0.04. The required bits per frame of the H264+RS-HARQ  increase from BER$=0.02$, where H264+RS(64, 127) cannot restore all the errors. When BER $> 0.04$, the  capability of the H264+RS-HARQ cannot correct  the errors and its performance decreases sharply. The SVC-HARQ first transmits 160 bits per frame. Then, 160 incremental  bits are transmitted if the proposed semantic detector finds errors. The SVC-HARQ can reach the best performance of SVC+RS(64, 127) when no bit error exists and becomes close to the SVC+RS(64 ,255) when BER $>$ 0.04. The AKD performance of the SVC-HARQ is not smooth when BER is between 0.02 and 0.06 because the error detection of detected AKD is not a strict method. 
The required bits per frame are the same as the  H264+RS(64, 127) when  no bit error exists and then increase with the BER.  Overall, the SVC-HARQ always requires 1/12 bits per frame of the H264+RS-HARQ. 

The absolute value changes of Ploss is small because the Ploss represents the whole content and  expression errors only occupy a small part in the image. For example, the Ploss of  the SVC-HARQ is about 0.18 when BER=0 and that is about 0.23 when BER =0.2. The AKD only concentrates on the facial expression and its change is large.  Specially, the tendencies of Ploss and AKD metrics are similar. Therefore, inaccurate facial expression is the  major factor for lower  Ploss due to higher BER.

From the above discussion, the SVC-HARQ shows its flexibility in the bit consumption with the change in BER. Meanwhile, the SVC-HARQ can reach the best performance of the SVCs trained under different BERs as an adaptive method. However, the semantic detector can only protect the fluency of the video. A detector to find expression errors should be proposed for an important conference.

\subsection{Performance of SVC-CSI }
We consider  frequency-selective channels to test the effectiveness of the CSI feedback. All the SVCs with CSI feedback are trained under the channels with exponential power delay profile of three paths. Each path obeys complex Gaussian distribution.  Because the channel model is introduced when training, we also simulate  untrained channels with  five paths to test the robustness of the SVC. The testing environment is called mismatched channel environments because their statistical parameters, such as delay spread, are different from trained channels. The number of the subchannels in the frequency domain is 16. The transmit bits are modulated to 16-QAM.

\begin{figure}[!h]
	\centering

\subfloat[ ]{ 
		\includegraphics[width=3.5in]{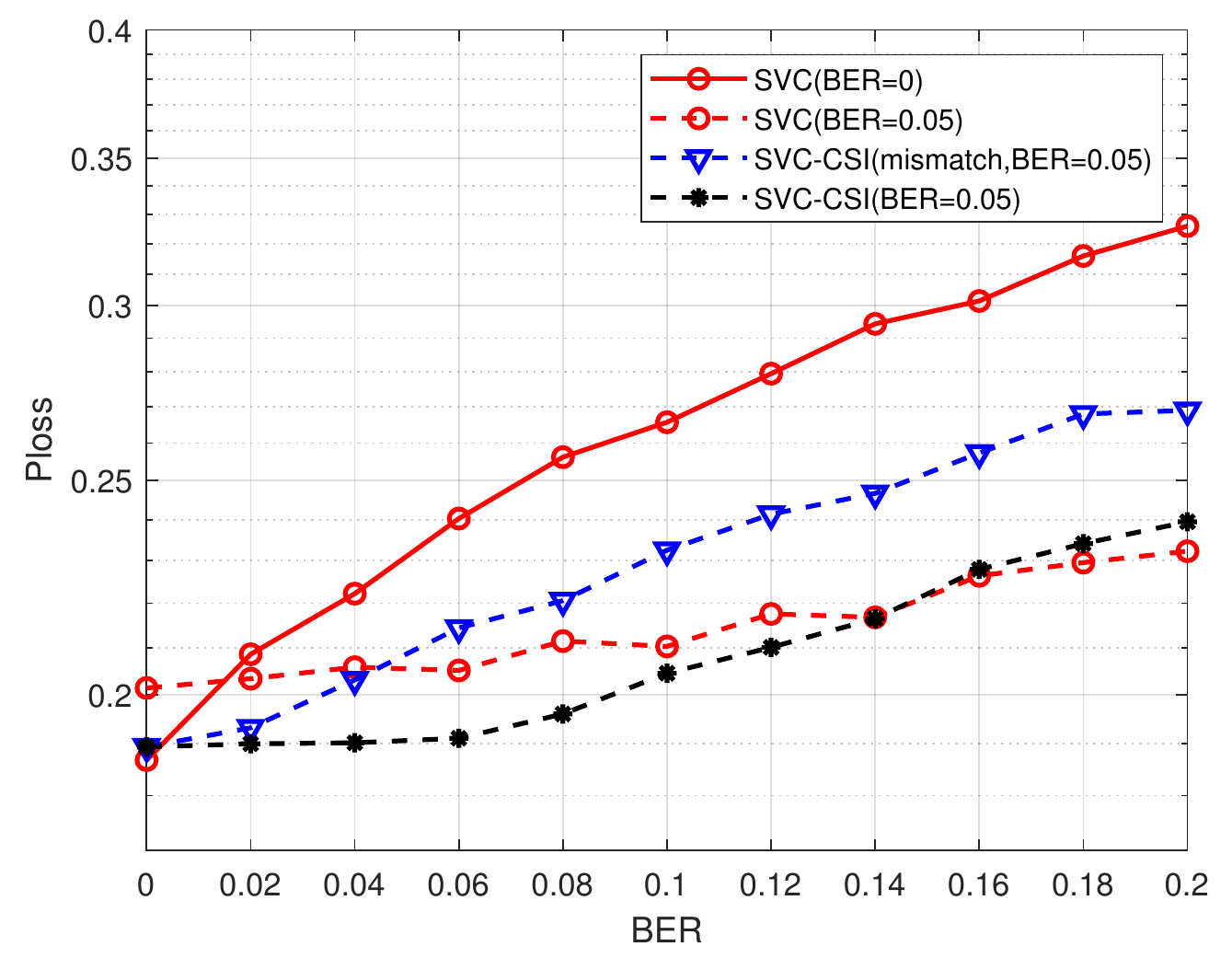}}
	\subfloat[ ]{ \includegraphics[width=2.5in]{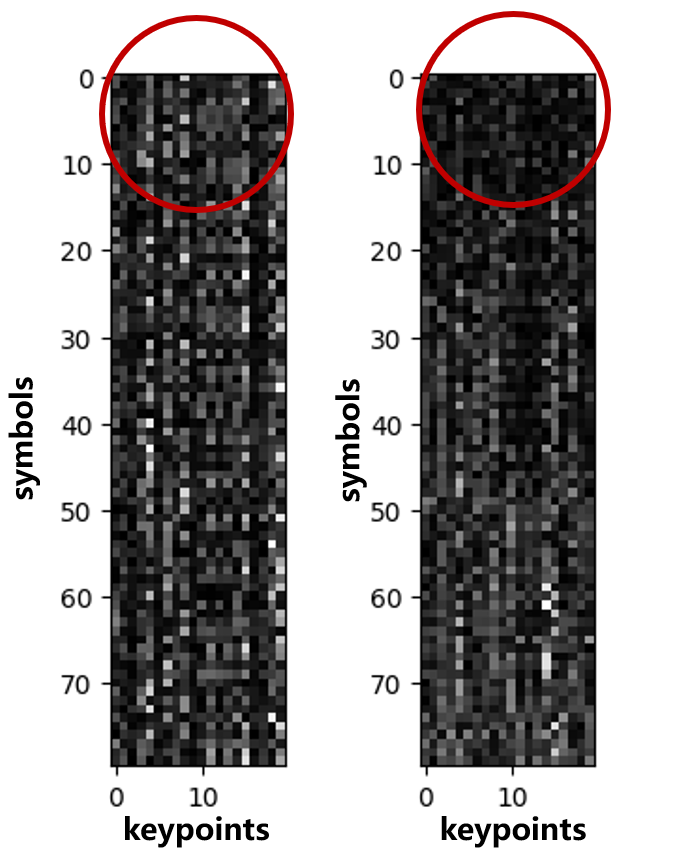}}

	% where an .eps filename suffix will be assumed under latex,
	% and a .pdf suffix will be assumed for pdflatex; or what has been declared
	% via \DeclareGraphicsExtensions.
	\caption{(a) Ploss performance of the SVC-CSI methods. (b) Gray image of the trained weights to visualize the effects of CSI feedback. The left image is trained without CSI information while the right one is trained under CSI feedback.  }
	\label{SVC_CSI}
\end{figure}
As shown in Fig. \ref{SVC_CSI}(a), the SVC-CSI(BER=0.05) means the average BER of the training channels is 0.05.    The SVC-CSI (BER=0.05) has similar Ploss performance as SVC (BER=0.05) when BER > 0.14, and has better performance than SVC (BER=0.05) when BER <0.14. It reaches the performance of SVC (BER=0) when BER=0 because the SVC-CSI learns to protect the information according to the qualities of the channels. However, the performance of the SVC-CSI decreases more sharply under mismatched channel environments and thus SVC-CSI (mismatch) performs  worse than the SVC (BER=0.05) when BER > 0.04. Overall, the CSI feedback enhances the performance when BER is low under the matched channel environments but loses its robustness under the mismatched environments.

In order to visualize the impact of the CSI feedback, we replace the three dense layers at the transmitter in the technical level with one dense layer, whose input includes  keypoints (20 real numbers) and output includes 80 symbols (320 bits with 4-bit quantization). In Fig. \ref{SVC_CSI}(b), the absolute values of trained multiplicative weights in this dense layer are   shown as a gray picture, and only the weights on the right picture are trained with CSI feedback.  Thus,
the $0\sim 10$ symbols on the right picture  are usually transmitted at  better channels than those on the left one due to CSI feedback. The absolute values in the circle of the right picture are larger than those of the left picture. This finding means that the transmitter  learns to place more information at the better channel  conditions.  The SVC-CSI performs better than the SVC when BER is higher because  most information is transmitted at the first several channels with lower noise power.

\begin{figure}[!h]
	\centering

	\subfloat[ ]{ 
		\includegraphics[width=2.5in]{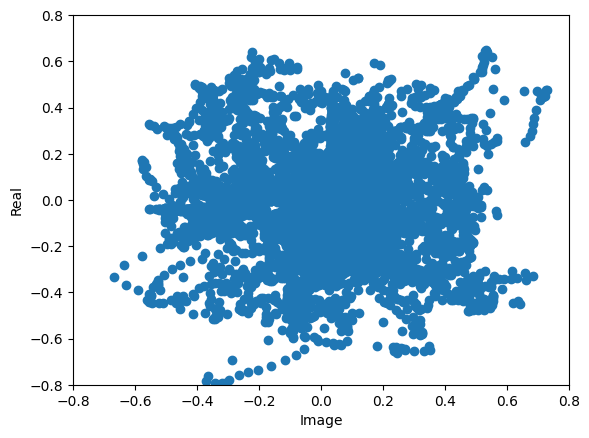}}
	\subfloat[ ]{ \includegraphics[width=2.5in]{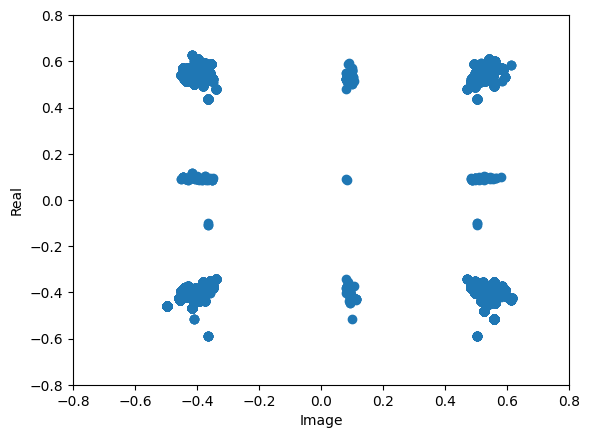}}\\
		\subfloat[ ]{ 
		\includegraphics[width=4in]{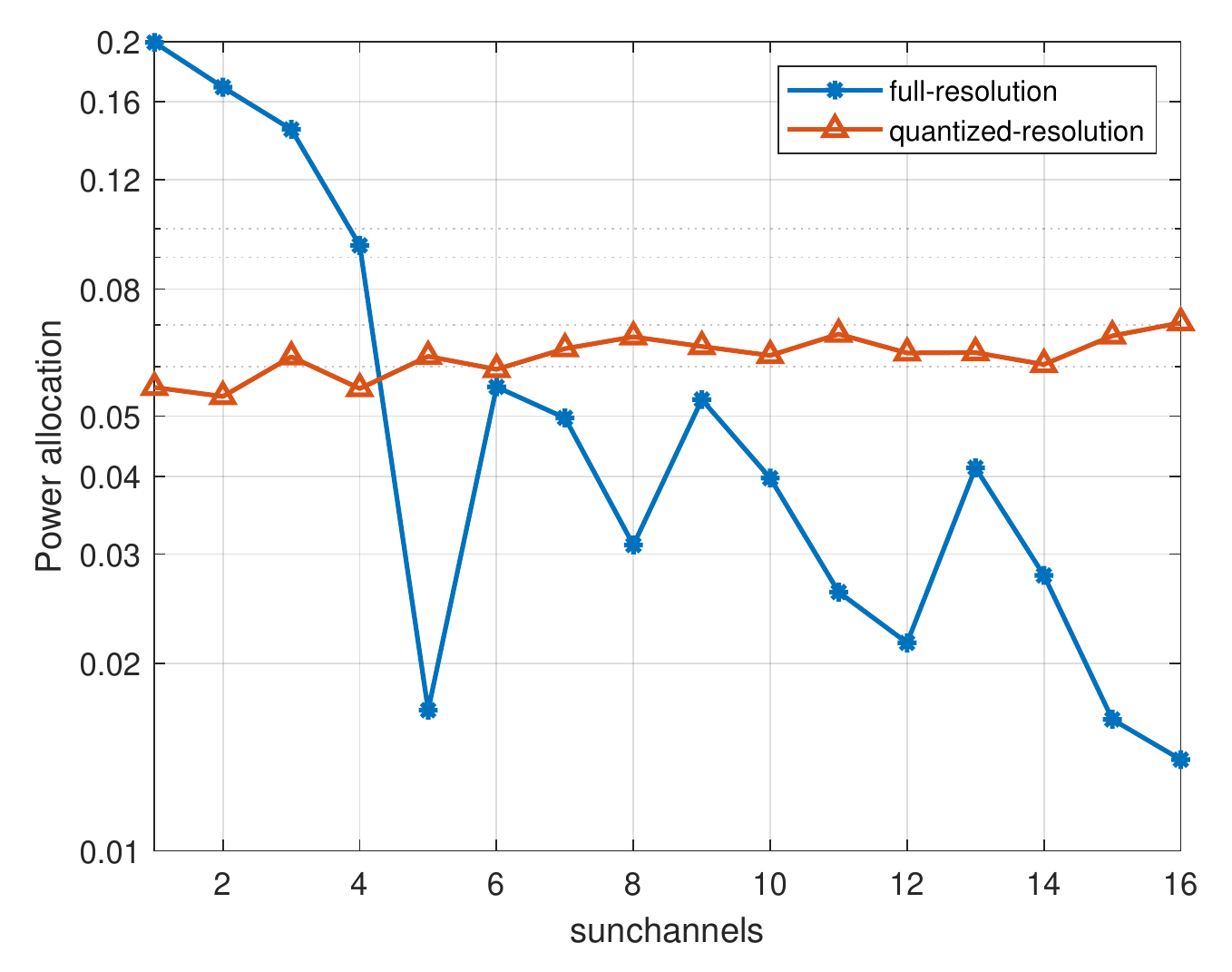}}

	% where an .eps filename suffix will be assumed under latex,
	% and a .pdf suffix will be assumed for pdflatex; or what has been declared
	% via \DeclareGraphicsExtensions.
	\caption{Different  constellations of the two SVC methods. (a) Constellation points of SVC-CSI (full-resolution). (b) Constellation points of SVC-CSI (quantized-resolution). (c) Different transmit powers  allocated at different subchannels. The channel condition becomes worse with the increase of the channel number. }
	\label{SVC_QAM}
\end{figure}

The keypoints are also transmitted as symbols directly under Rayleigh fading channels in Fig. \ref{SVC_QAM}.  The SVC-CSI (full-resolution) learns to transmit 160 real symbols. Due to CSI feedback with 16 subchannels, the first five symbols are always transmitted under the highest SNR and the last five symbols are under the lowest SNR.   The constellation points of SVC-CSI (full-resolution) are  spread around in Fig. \ref{SVC_QAM}(a). Meanwhile, this method learns to transmit more information at better subchannels; thus, the transmit power decreases when the channel condition becomes worse as shown in Fig. \ref{SVC_QAM}(c). The SVC-CSI (quantized-resolution) has the same modulation method at all subchannels and its constellation points shown in Fig. \ref{SVC_QAM}(b) are similar to 16-QAM. In order to cope with the noise, the transmit power of the SVC-CSI (quantized-resolution) becomes larger as the channel condition becomes worse. 

\begin{figure}[!h]
	\centering

		\includegraphics[width=4in]{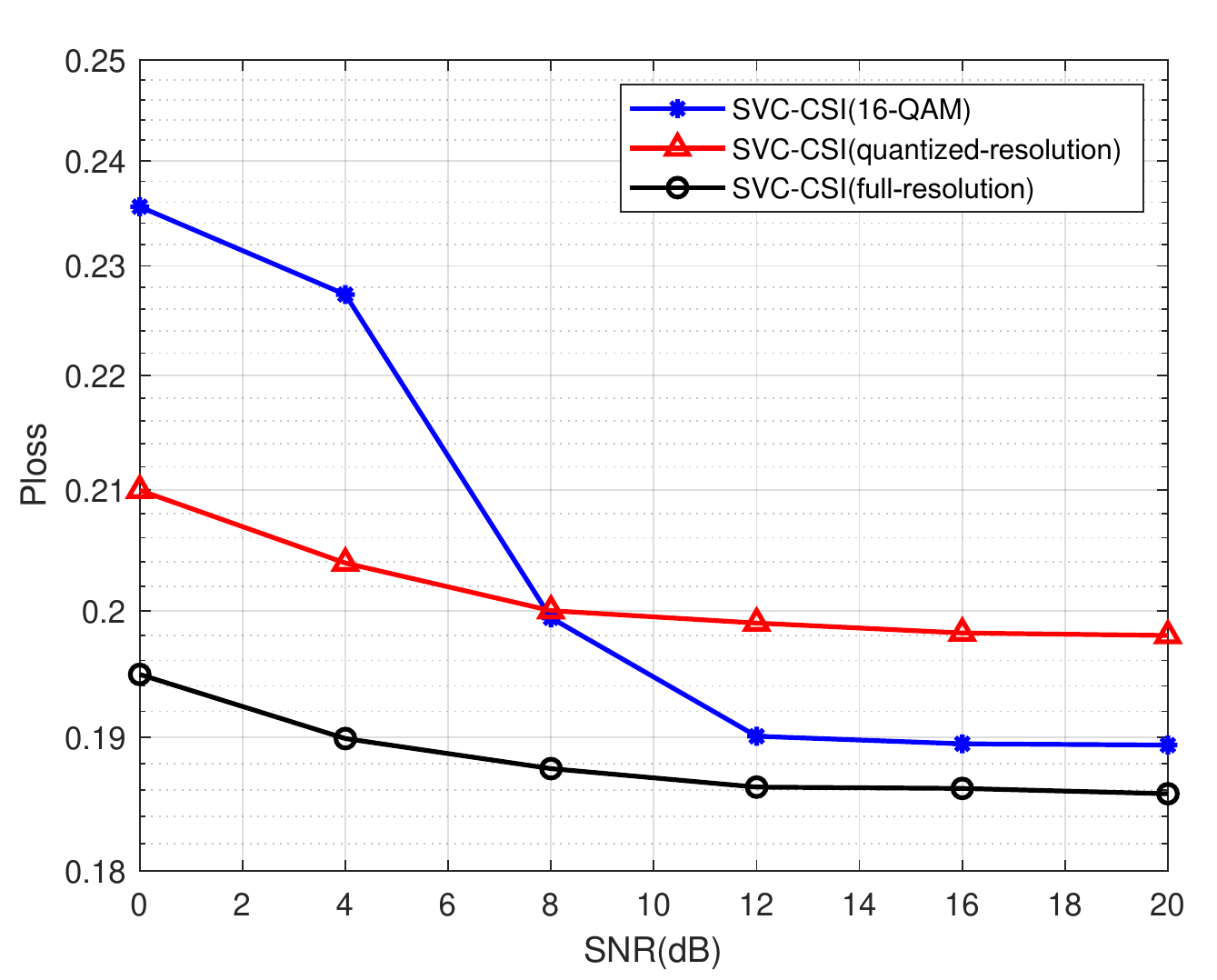}

	% where an .eps filename suffix will be assumed under latex,
	% and a .pdf suffix will be assumed for pdflatex; or what has been declared
	% via \DeclareGraphicsExtensions.
	\caption{Performance of different constellation modes.  }
	\label{SVC_AWGN}
\end{figure}

The Ploss performance of the SVC-CSI under fading channels is compared. The SVC-CSI (16-QAM) is trained to transmit 320 bits and modulated to 80 16-QAM symbols. The three methods in Fig. \ref{SVC_AWGN} have the same transmit resources. The SVC-CSI (full-resolution) always  has best performance, but its complexity is impractical. The SVC-CSI (quantized-resolution) learns a different modulation method from 16-QAM.  This method performs  worse than 16-QAM when SNR $\geq$ 8 dB but better than 16-QAM when SNR is low. Therefore, the trained modulation of  SVC-CSI (quantized-resolution)  is suitable for wired environments but cannot perfectly reconstruct the frame when SNR is high. 

  	\begin{table}[!h]
 	\centering	
 	\caption{Ploss performance of SVC-CSI-HARQ under matched and mismatched channels. }
 	\begin{tabular}{>{\sf }c|cccccc}    %
 		\toprule
 		BER	& 0 & 0.02&0.04&0.06&0.08&0.10\\ \hline
 		SVC-HARQ	&\bf{0.186}&	0.199	&0.204	&0.205	&0.203&	0.207 \\ \hline
 		
 		SVC-CSI-HARQ(mismatch)& 0.189 &	0.193&	0.203&	0.205&	0.206	&0.207\\ \hline
 		SVC-CSI-HARQ &0.189&	\bf{0.190}&	\bf{0.190}&	\bf{0.191}&	\bf{0.196}&	\bf{0.206}\\ 
 		
 		\bottomrule
 	\end{tabular}
 	\label{T1}
 \end{table}

The introduction of HARQ can improve the performance of the SVC-CSI under mismatched channels, and this method is called SVC-CSI-HARQ. In this method, the second transmission is trained without CSI feedback and thus robust to the varying environments. This strategy can guarantee that the SVC-CSI-HARQ under mismatched environment is not worse than SVC-HARQ. Meanwhile, SVC-CSI-HARQ performs better than SVC-HARQ under mismatched channels when 0.02 $\leq$ BER $\leq$ 0.1 because the mismatched  power correlation at subchannels is slight when subchannel gain is large. In addition, the SVC-CSI-HARQ shows its superiority under the matched channels.    Specially, the SVC-HARQ is slightly better than the methods with CSI feedback when BER=0 because the CSI feedback is  ineffective under noiseless channels.   In contrast, the CSI feedback may mislead the  SVC to transmit less information at the last subchannels

\section{Conclusions}
We have investigated  semantic transmission framework for video conferencing. The knowledge of the speaker's photos can be shared explicitly because the faces play an essential role at a conference.  The three-level framework, called SVC, is established by utilizing only keypoint transmission to represent the motion of the facial expressions.  The compression by the SVC only loses  detailed expressions while the conventional methods reduce the resolution. Furthermore, the transmission errors in the conventional methods destroy pixels directly  while those in SVC leads  a changed expression.

We have also considered the impact of feedback in the SVC and designed an IR-HARQ framework called SVC-HARQ with ACK feedback. The changed  expression of the error keypoints obtains nonsmooth adjacent frames. To detect semantic errors, we have developed a semantic detector, including identity classifier and fluency detection. The SVC-HARQ is  flexible and it can combine the performance of the networks trained under different BERs and always reach a good performance. 
The CSI feedback can also enhance the performance further. The transmitted symbols or bits are sorted by SNRs at different subchannels, called SVC-CSI.  The SVC-CSI learns to allocate more information at the subchannels with higher gains and performs better than the SVC without CSI feedback.   However, the robustness of the SVC-CSI decreases because the channel model is exploited when training. The combination of CSI and ACK feedback can balance the performance, bit consumption, and robustness.

	\bibliographystyle{IEEEtran}
	\bibliography{bibtex0320}
	
	
	%%%%\begin{IEEEbiography}{Michael Shell}
	%%%%Biography text here.
	%%%%\end{IEEEbiography}
	%%%%
	%%%%% if you will not have a photo at all:
	%%%%\begin{IEEEbiographynophoto}{John Doe}
	%%%%Biography text here.
	%%%%\end{IEEEbiographynophoto}
	%%%%
	%%%%% insert where needed to balance the two columns on the last page with
	%%%%% biographies
	%%%%%\newpage
	%%%%
	%%%%\begin{IEEEbiographynophoto}{Jane Doe}
	%%%%Biography text here.
	%%%%\end{IEEEbiographynophoto}
	
	
	%\vfill
	

\end{document}